\documentclass[twocolumn]{aastex631}

\usepackage{amsmath}
\usepackage{amssymb}
\usepackage{amsthm}
\usepackage{multirow}
\usepackage{mathrsfs}

\begin{document}

\title{
%\sout{Differential Emission Measure Analysis of the X-ray Gas in SN 1987A from 2007 to 2021: The Fading Ring and the Brightening Ejecta}\\
Evolution of X-ray Gas in SN 1987A from 2007 to 2021: Ring Fading and Ejecta Brightening Unveiled through Differential Emission Measure Analysis\\
%%%\lsun{Any suggestions for a better title?}
}

	\correspondingauthor{Lei Sun}
	\email{l.sun@nju.edu.cn}
	
	\author[0000-0001-9671-905X]{Lei Sun}
	\affiliation{Department of Astronomy, Nanjing University, Nanjing 210023, People's Republic of China}
	\affiliation{Key Laboratory of Modern Astronomy and Astrophysics, Nanjing University, Ministry of Education, People's Republic of China}
    \affiliation{Anton Pannekoek Institute \& GRAPPA, University of Amsterdam, PO Box 94249, 1090 GE Amsterdam, The Netherlands}

    \author[0000-0003-2836-540X]{Salvatore Orlando}
    \affiliation{INAF-Osservatorio Astronomico di Palermo, Piazza del Parlamento 1, 90134 Palermo, Italy}

    \author[0000-0001-5792-0690]{Emanuele Greco}
    \affiliation{Anton Pannekoek Institute \& GRAPPA, University of Amsterdam, PO Box 94249, 1090 GE Amsterdam, The Netherlands}
    \affiliation{INAF-Osservatorio Astronomico di Palermo, Piazza del Parlamento 1, 90134 Palermo, Italy}
    \affiliation{Universit\`a degli Studi di Palermo, Dipartimento di Fisica e Chimica E. Segr\`e, Piazza del Parlamento 1, 90134 Palermo, Italy}

    \author[0000-0003-0876-8391]{Marco Miceli}
    \affiliation{Universit\`a degli Studi di Palermo, Dipartimento di Fisica e Chimica E. Segr\`e, Piazza del Parlamento 1, 90134 Palermo, Italy}
    \affiliation{INAF-Osservatorio Astronomico di Palermo, Piazza del Parlamento 1, 90134 Palermo, Italy}

    \author[0000-0002-4753-2798]{Yang Chen}
    \affiliation{Department of Astronomy, Nanjing University, Nanjing 210023, People's Republic of China}
    \affiliation{Key Laboratory of Modern Astronomy and Astrophysics, Nanjing University, Ministry of Education, People's Republic of China}

    \author[0000-0002-4708-4219]{Jacco Vink}
    \affiliation{Anton Pannekoek Institute \& GRAPPA, University of Amsterdam, PO Box 94249, 1090 GE Amsterdam, The Netherlands}

    \author[0000-0002-5683-822X]{Ping Zhou}
    \affiliation{Department of Astronomy, Nanjing University, Nanjing 210023, People's Republic of China}
	\affiliation{Key Laboratory of Modern Astronomy and Astrophysics, Nanjing University, Ministry of Education, People's Republic of China}



\begin{abstract}

%%%\lsun{to be finished}
{As the nearest supernova (SN) observed since Kepler's SN of 1604, SN 1987A provides an unprecedented opportunity to study in detail the early evolution of supernova remnants (SNRs). Despite extensive studies through both observations and simulations, there is still an urgent need for a more effective approach to integrate the results from two sides. In this study, we conducted a detailed differential emission measure (DEM) analysis on the XMM-Newton observations taken in 2007 to 2021 to characterize the continuous temperature structure of SN 1987A, which can be better compared with simulations. The X-ray plasma exhibit a temperature distribution with a major peak at $\sim0.5$--$1$\,keV and a high-temperature tail extending to $\gtrsim5$\,keV. The emission measure (EM) of the major peak started to decline around 2014, while the EM of the tail continued increasing and appears to have formed a secondary peak at $\sim3$--$5$\,keV in recent years. Our DEM results consistent well with simulations, which help to further identify the major peak as originating from the equatorial ring and the secondary peak as arising from the newly shocked ejecta. Together with the simulations, our DEM analysis reveals recent fading of the ring and brightening of the ejecta in X-rays from SN 1987A. Additionally, we observed a recent decrease in the centroid energy of Fe K line, providing further evidence of newly shocked ejecta.}

\end{abstract}



\section{Introduction}

%%%\lsun{to be finished}
{The supernova remnant (SNR) is the leftover of a supernova (SN) explosion, formed by the interaction of the SN ejecta with the ambient circumstellar material (CSM) and interstellar medium (ISM). Therefore, it provides a crucial perspective to study the physical and chemical properties of both the ejecta and the CSM/ISM. These studies can help us gain a better understanding of SNR-related physics, such as remnant evolution and its {metal}/energy feedback, collisionless shock physics, cosmic-ray acceleration mechanism, etc. Additionally, they allow us to trace back the progenitor properties and explosion mechanisms of SNe.

SN 1987A was a Type II SN exploded in the Large Magellanic Cloud (LMC), {whose progenitor is a blue supergiant \citep[BSG,][]{1987A&A...177L...1W},} and was first detected on 1987 February 23. As the nearest SN observed since Kepler's SN of 1604, SN 1987A provides an unprecedented opportunity to study in detail the SN-SNR connection and the early phase of SNR evolution. The remnant of SN 1987A is represented by its well-known triple-ring CSM system, where one equatorial ring (ER) together with two outer rings constitute an hourglass-like structure. Starting from $\sim4000$ days after the explosion, several ``hot spots'' emerged in optical band as a result of the SN blast wave (i.e., the forward shock) encountering the dense inner layer of ER \citep[e.g.,][]{1998ApJ...492L.139S,2000ApJ...537L.123L}. Soon after, the main body of the ER began to be impacted and heated by the forward shock, leading to a dramatic increase in its soft X-ray brightness at $\sim6000$ days \citep[e.g.,][]{2005ApJ...634L..73P}. From then on, the blast wave continuously interacted with the ER, made it shining across all the energy bands. Until recently, the forward shock seems to have traversed the main ER but still propagating into the high-latitude materials, indicated by the decreasing optical, infrared, and soft X-ray fluxes, and the still increasing hard X-ray flux \citep[e.g.,][]{2015ApJ...806L..19F,2016AJ....151...62A,2019ApJ...886..147L,2021ApJ...916...41S,2024ApJ...966..147R}. {On the other hand, the evidence of a pulsar wind nebula (or a cooling neutron star) at the center of SN 1987A has recently be found based on the NuSTAR detection of its hard-band non-thermal emission \citep{2022ApJ...931..132G} and the JWST detection of emission lines due to its ionizing radiation \citep{2024Sci...383..898F}.} 

SN 1987A is also one of the few SNRs {that have been studied in great detail through} three-dimensional (3D) hydrodynamic (HD) and magneto-hydrodynamic (MHD) simulations tracing their long-term evolution from the explosion all the way to the remnant phase \citep[e.g.,][]{2015ApJ...810..168O,2019A&A...622A..73O,2020A&A...636A..22O,2020ApJ...888..111O}. The simulations have successfully reproduced the multiwavelength emission of SN 1987A, where the synthetic images, light curves and spectra are well comparable with the observations up to now. In particular, the synthetic high-resolution X-ray spectra show an impressive consistency with the {the Chandra HETG \citep{2019NatAs...3..236M} and XMM-Newton RGS \citep{2022ApJ...931..132G} observations at multiple epochs.}

Despite extensive study of SN 1987A through both observations and simulations, several unresolved issues persist. Specifically, there is still a need for a more effective approach to connect observations with simulations. Most previous studies on SN 1987A adopted discrete-temperature (either two-temperatures or three-temperatures) plasma models to fit its X-ray spectra \citep[e.g.,][]{2006A&A...460..811H,2004ApJ...610..275P,2006ApJ...646.1001P,2006ApJ...645..293Z,2009ApJ...692.1190Z,2010MNRAS.407.1157Z,2008ApJ...676..361H,2010A&A...515A...5S,2012ApJ...752..103D,2016ApJ...829...40F,2020ApJ...899...21B,2021ApJ...916...76A,2021ApJ...916...41S,2021ApJ...922..140R,2021ApJ...908L..45G,2022ApJ...931..132G,2022A&A...661A..30M,2024ApJ...966..147R}. As summarized by \citet[][see Table 6 therein]{2021ApJ...916...76A}, these works found plasma temperatures in the range of $\sim0.5$--$4$\,keV, with a low-temperature component at $\sim0.5$\,keV, a high-temperature component at $\sim2$--$4$\,keV, and some of them a third intermediate-temperature component at $\sim1$\,keV. However, while these discrete-temperature models can approximately fit the spectra and characterize the average properties of the plasma, they may be inadequate to fully represent the complex structure of SN 1987A and are difficult to be directly compared with the continuous parameter distributions evaluated from simulations. More importantly, interpretations based on these modeling might also be biased.

This study aims to characterize the continuous temperature structure of the X-ray gas in SN 1987A through a detailed differential emission measure (DEM) analysis of the long-term XMM-Newton observations. The results of the DEM analysis will be compared with MHD simulation predictions, which will in turn provide a more comprehensive interpretation of the observations and enhance our understanding of remnant evolution. Similar attempts have been made by \citet{2006ApJ...645..293Z,2009ApJ...692.1190Z} using Chandra LETG/HETG data and by \citet{2021ApJ...916...76A} using RGS $+$ NuSTAR data. However, their results are not consistent with each other, and neither of them agree well with the simulations, which motivated us to perform further investigations into this issue. 

We describe the observations and the data reduction procedures in Section \ref{sec:obs}, describe the DEM model and present the DEM analysis results in Section \ref{sec:DEM}, discuss the DEM results in Section \ref{sec:discussion} by comparing observations with simulations, and make our conclusions in Section \ref{sec:conclusion}.}

\section{Observations and Data Reduction}\label{sec:obs}

SN 1987A has been regularly monitored in X-rays. In this work, we utilized a set of XMM-Newton observations taken since 2007, which is similar to those used in \citet{2021ApJ...916...41S}, but with {two additional epochs: the first was taken in November 2020 (PI: F.~Haberl), and the second was obtained in} December 2021 (PI: L.~Sun), representing the most recent {XMM-Newton observation of the} evolution of SN 1987A. For each observation, we used both the RGS and the EPIC-pn exposures. The high energy resolution of RGS can help to detect and resolve the numerous emission lines in $0.35$--$2.5$\,keV, which is crucial for our spectral modeling. EPIC-pn has a large effective area and a complete energy coverage up to 10\,keV, providing further constraints on the continuum and the high-temperature plasmas. The details of all observations are summarized in Table \ref{tab:obs}.

All the data were processed using the XMM-Newton Science Analysis Software (SAS, version 18.0.0)\footnote{https://www.cosmos.esa.int/web/xmm-newton/sas} with the latest calibration files. We extracted the RGS and the EPIC-pn spectra following {the procedures {described} in} \citet{2021ApJ...916...41S}. The spectra were then optimally rebinned adopting the \citet{2016A&A...587A.151K} optimal binning scheme.

\begin{deluxetable*}{ccccccc}
	\tablecaption{Observations\label{tab:obs}}
	\tablenum{1}
	\tablehead{
		\multicolumn{3}{c}{}&\multicolumn{2}{c}{$t_{\rm exp}$ (ks)\tablenotemark{a}}&\multicolumn{2}{c}{$\sum$GTI (ks)\tablenotemark{b}}\\
		ObsID & Date&Age (days) & EPIC-pn & RGS & EPIC-pn & RGS
        }
	\startdata
	    0406840301 & 2007 Jan 17 & 7267 & 106.9 & 111.3 & 61.1 & 109.8 \\
	    0506220101 & 2008 Jan 11 & 7627 & 110.1 & 114.3 & 70.7 & 102.4 \\
		0556350101 & 2009 Jan 30 & 8012 & 100.0 & 101.9 & 66.4 & 101.8 \\
		0601200101 & 2009 Dec 11 & 8327 & 89.9 & 91.8 & 82.4 & 91.7 \\
		0650420101 & 2010 Dec 12 & 8693 & 64.0 & 65.9 & 52.7 & 65.9 \\
		0671080101 & 2011 Dec 2 & 9048 & 80.6 & 82.5 & 64.2 & 80.6 \\
		0690510101 & 2012 Dec 11 & 9423 & 68.0 & 69.9 & 59.4 & 69.8 \\
		0743790101 & 2014 Nov 29 & 10141 & 78.0 & 79.6 & 56.4 & 79.4 \\
		0763620101 & 2015 Nov 15 & 10492 & 64.0 & 65.9 & 58.0 & 65.8 \\
		0783250201 & 2016 Nov 2 & 10845 & 72.4 & 74.3 & 50.3 & 74.2 \\
		0804980201 & 2017 Oct 15 & 11192 & 77.5 & 79.4 & 27.8 & 79.3 \\
		0831810101 & 2019 Nov 27 & 11964 & 32.4 & 34.9 & 11.1 & 34.8 \\
        0862920201 & 2020 Nov 24 & 12328 & 77.6 & 79.7 & 57.8 & 72.8 \\
        0884210101 & 2021 Dec 28 & 12727 & 90.6 & 89.4 & 75.2 & 86.1 \\
	\enddata
	\tablenotetext{a}{Total exposure times.}
	\tablenotetext{b}{Total good time intervals after background flare removal.}
\end{deluxetable*}

\section{Differential Emission Measure Analysis}\label{sec:DEM}

In this section, we describe the DEM model, verify the validity and the reliability of the model, and then present the DEM analysis results of SN 1987A. We used XSPEC (version 12.12.1)\footnote{https://heasarc.gsfc.nasa.gov/xanadu/xspec/} with AtomDB 3.0.9\footnote{http://www.atomdb.org/} for spectral analysis, and adopted C statistic. Unless otherwise specified, in this paper the metal abundances are with respect to their solar values \citep{2000ApJ...542..914W}, and the error bars represent the 1-$\sigma$ uncertainties.

\subsection{Differential Emission Measure Model}\label{sec:DEM_model}

Our DEM model is similar to those used in \citet{2006ApJ...645..293Z,2009ApJ...692.1190Z} and \citet{2021ApJ...916...76A}, which is basically an extended version of the XSPEC model {\tt c6pvmkl} \citep{1989ApJ...341..474L,1996ApJ...456..766S} that works not only on the collisional-ionization-equilibrium (CIE) plasmas but also on the non-equilibrium-ionization (NEI) plasmas.

{The {volume} emission measure (EM) of a {thermally-emitting} plasma, defined as $EM\equiv\int n_{\rm e}n_{\rm H}dV$, can be described by a function of temperature as:
\begin{equation}\label{eq:DEM}
    EM=\int d(EM)=\int \phi(T)dT=\int T\phi(T)d(\ln T)
\end{equation}
where $\phi(T)$ is called the DEM function, representing the temperature distribution of the plasma.
In our model,} the DEM function $\phi(T)$ is parameterized by an $M^{\rm th}$-order Chebyshev series as:
\begin{equation}\label{eq:chebyshev}
    \phi(T)=\alpha e^{\omega(T)},\quad\omega(T)=\sum^M_{k=1}a_kP_k(T)
\end{equation}
where $\alpha$ is the normalization parameter, $a_k$ the coefficients, {$P_k$ the Chebyshev polynomial of order $k$, and $M$ the maximum order of the Chebyshev series}.
$T$ denotes the (quasi-)continuous electron temperature, which is represented by $N$ temperature bins ($T_i$, $i=1,2,...,N$) logarithmically spaced in 0.1--10\,keV. For each $T_i$, the X-ray spectrum is calculated by a single {\tt vnei} model, with $T_e=T_i$ and {the} {EM} given by $\phi(T_i)$. In order to lower the number of free parameters and to keep the model as simple as possible, we assume that the ionization parameter $\tau_i=(n_et)_i$, {where $t$ denotes the elapsed time since the gas was shocked}, is a power-law function of the temperature following \citet{2006ApJ...645..293Z,2009ApJ...692.1190Z} and \citet{2021ApJ...916...76A}, i.e.,
\begin{equation}\label{eq:net}
    \tau_i=\tau_{\rm 1\,keV}\left(\frac{T_i}{\rm 1\,keV}\right)^\beta
\end{equation}
where $\tau_{\rm 1\,keV}$ is the ionization parameter at 1\,keV, {and} $\beta$ the power-law index. {Despite that the X-ray emitting plasma may have a non-uniform complex chemical composition, we assume {that its metal abundances are identical across all} temperature bins. {This assumption is viable, given that the majority of the X-ray emission originates from the shocked CSM and from the shocked outermost layers of ejecta (expected with abundances similar to those of CSM), and thus only small variations of abundances are expected from the different plasma components.} Following \citet{2010A&A...515A...5S} and the references therein, with abundances converted to those in \citet{2000ApJ...542..914W}, the abundances of N, O, Ne, Mg, Si, S, and Fe are set as free parameters, while those of He (set to 2.57), C (0.14), Ar (0.76), Ca (0.49), and Ni(0.98) are fixed.} %%\lsun{[response to EG ``- {\it It is not clear to me how the ionization parameter and the abundances values are included in the Eq 1. it would be helpful to explicit mention it since we refer to the vnei model which also relies on these parameters.}'': I have moved the descriptions on the abundance setup to here.]} 
Then the final spectrum is the sum of all the spectra in all the temperature bins (therefore, the DEM model could also be {regarded} as a combination of $N$ {\tt vnei} components).

%\lsun{response to EG and SO: I absolutely agree with Emanuele that the power-law assumption of ionization parameter may lead to some bias especially at high temperatures. But as you and Salvo mentioned, it is very difficult to have an even more complicated description of Tau, which may induce more problems. I have added some further discussions on this in Section 4.3, in order to warning the readers on possible effects.}

The maximum order of the Chebyshev series $M$ and the total temperature bin number $N$ may affect the modeling performance and the fitting results, therefore should be chosen with caution based on the data fitted. A rather large $M$ and $N$ can help to achieve a high temperature resolution and to capture delicate structures in DEM distribution, but will introduce more free parameters, consume a lot of computer time, and may lead to over-fitting when {applied} to low-quality data. On the other hand, a rather small $M$ and $N$ may be not enough to reproduce the DEM distribution and {lose} information in the data. We have tried different combinations of $M$ and $N$ in our test cases {(see Appendix \ref{app:M_N} for details)}, and finally fix them as $M=7$ and $N=40$, which achieves the best performance for our RGS $+$ EPIC-pn spectral fitting {of SN 1987A}. Following the naming logic of {\tt c6pvmkl} model, we call this DEM model as {\tt c7pvnei} hereafter. %%\lsun{[response to YC ``{\it - whether this model is generic or is specific designed for SN 1987A?}'': The DEM model is constructed for a general use, while the detailed parameter setup, together with $M$ and $N$, are specifically optimized for RGS + EPIC-pn spectra of SN 1987A. However, all these parameters, including $M$ and $N$, are variable in this model, thus can be applied to any other observations.]}

Apart from the DEM component, the overall setup of the spectral model is similar to those in \citet{2021ApJ...916...41S}. The X-ray emission from SN 1987A is subjected to absorption from both the Galactic and the LMC ISM, which are characterized by two {\tt tbvarabs} components. The Galactic absorption is fixed at $N_{\rm H,Gal}=6\times10^{20}$\,{cm$^{-2}$} with solar abundances, while the LMC absorption $N_{\rm H,LMC}$ is set as free parameter ({see the discussions below}) with the average LMC abundances given by \cite{1992ApJ...384..508R}. A {\tt gsmooth} component and a {\tt vashift} component are involved to account for the line broadening effect and the Doppler shift, respectively, and their parameters are fixed to the best-fit values given in \citet{2021ApJ...916...41S} {(see the values in Table 2 therein)}.

{The LMC absorption $N_{\rm H,LMC}$ is not tied between different observations due to two major reasons. Firstly, from a physical perspective, while it is improbable that there would be varying large-scale ISM absorption, there remains the possibility of an evolving local absorption component arising from both the unshocked CSM and the unshocked ejecta of SN 1987A. As pointed out by \citet{2021ApJ...916...41S}, a density of $1\times10^4$\,H\,cm$^{-3}$ and a path length of 0.01\,pc will result in a measurable local absorption $N_{\rm H}\sim0.3\times10^{21}$\,cm$^{-2}$. This local absorption may change over time due to the expansion of the ejecta, potential destruction of CSM clumps, and rapid changes in the X-ray surface brightness distribution of the remnant itself \citep[see the Chandra images in][]{2016ApJ...829...40F}. Secondly, from a technical standpoint, it is nearly impossible to simultaneously fit all 14 observations with $N_{\rm H,LMC}$ tied up, since it will introduce too many free parameters and consume significant computational resources. By setting $N_{\rm H}$ as free parameter, we obtained a column density varying between $\sim2.2$--$2.8\times10^{21}$\,cm$^{-2}$, with a mean value $\sim2.5\times10^{21}$, cm$^{-2}$, which is consistent with the local absorption scenario. {In the time interval we have studied, the absorption column density was found to be increasing in general with some fluctuations. One possible explanation could be, as the ejecta expanding, more and more un-shocked ejecta material will be projected onto the ER region, and thus increase the total absorption column density. But on the other hand, the expansion of the ejecta will result in a decrease in its density, thereby we may expect to see a decrease in the absorption column density at some point in the future.} A further discussion on the local absorption will be interesting, but may be beyond the scope of this paper.}

As a brief summary, our spectral model can be described as: ${\tt tbvarabs}_{\rm Gal}\times{\tt tbvarabs}_{\rm LMC}\times{\tt gsmooth}\times{\tt vashift}\times{\tt c7pvnei}$. There are 18 free parameters in total: the LMC absorption column density ($N_{\rm H,LMC}$), 7 Chebyshev polynomial coefficients ($a_k,\ k=1,2,...,7$), 7 metal abundances (N, O, Ne, Mg, Si, S, and Fe), the ionization parameter at 1\,keV ($\tau_{\rm 1\,keV}$), the ionization parameter index ($\beta$), and finally the normalization ($\alpha$). As a comparison, a three-temperature model (3-T {\tt vnei} or 3-T {\tt vpshock}) with a similar parameter setup \citep[e.g.,][]{2021ApJ...916...76A,2022ApJ...931..132G} will have $\sim17$ free parameters. Thereby, the DEM model, with only one more free parameter than the traditional discrete-temperature models, is able to characterize the (quasi-)continuous distribution of the hot plasma among a {wide range in temperature space}, which provides us an unique insight into the real thermal condition of the gas and the underlying physics. In view of the large free parameter number and the potential degeneracy between different parameters, we ran a Markov Chain Monte Carlo (MCMC) simulation using the XSPEC {\tt chain} command after every fit, in order to determine the 1-$\sigma$ errors. The MCMC {simulation} was running using the Goodman-Weare algorithm with 40 walkers and a total length of $10^5$.

\subsection{Test on Simulated Data}\label{sec:test}

Before applying the DEM model to the real RGS $+$ EPIC-pn data, it is necessary to test the validity and the reliability of the model. At least, we need to ensure that if the X-ray gas in SN 1987A indeed has a bimodal or trimodal temperature distribution as previous studies suggested \citep[e.g.,][]{2006ApJ...645..293Z,2009ApJ...692.1190Z,2021ApJ...916...76A}, the DEM model should be able to reproduce those features.

\begin{deluxetable}{ccc}
    \tablecaption{Parameters of the test models\label{tab:test_model}}
	\tablenum{2}
	\tablehead{
             & $kT_e$ (keV) & $n_et$ ($10^{11}$\,s\,cm$^{-3}$)
            }
    \startdata
        1-T {\tt vnei} & 0.8 & 1.0 \\
        \hline
        \multirow{2}{4em}{2-T {\tt vnei}}& 0.5 & 10.0 \\
         & 2.5 & 1.0 \\
        \hline
        \multirow{3}{4em}{3-T {\tt vnei}}& 0.3 & 10.0 \\
         & 1.0 & 3.4 \\
         & 4.0 & 1.0 \\
    \enddata
\end{deluxetable}

\begin{figure}[ht]
    \centering
    \includegraphics[width=0.793\linewidth]{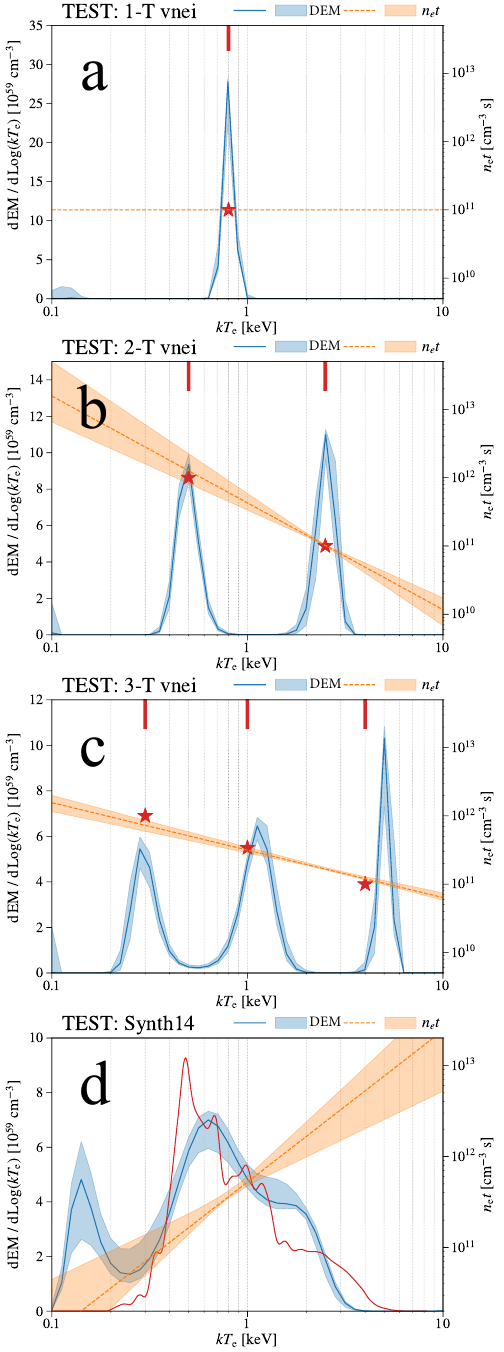}
    \caption{The DEM model {applied} on simulated spectra. Panel a--c shows the DEM fitting results of the 1-T, 2-T, and 3-T {\tt vnei} spectra, respectively. The red lines and stars {denote} the temperatures and ionization parameters used for simulating the spectra, the blue lines and areas show the best-fit DEM distributions and their 1-$\sigma$ uncertainties, and the orange dashed lines and areas show the best-fit $n_et$ distributions and their 1-$\sigma$ uncertainties. Panel d shows the fitting results of the synthetic spectra representing the simulated X-ray emission of SN 1987A in 2014 \citep{2022ApJ...931..132G}. The red line denotes the synthetic DEM distribution.}
    \label{fig:test}
\end{figure}

Therefore we constructed three test models that consist of one, two, and three {\tt vnei} components, respectively. The electron temperatures and the ionization parameters of the {\tt vnei} components in each test model are listed in Table \ref{tab:test_model}. The {EMs} of the {\tt vnei} components were set to the values that give a similar flux as SN 1987A \citep[$\sim7.8\times10^{-12}$\,erg\,cm$^{-2}$\,s$^{-1}$ in 0.5-2\,keV band at its peak level around 2013, see, e.g.,][]{2021ApJ...916...41S}. Other parameters, such as the absorption column density and metal abundances, were set following the average values obtained in \citet{2021ApJ...916...41S}. We then simulated the 80\,ks RGS (including the 1st and 2nd order spectra of RGS1 and RGS2) and EPIC-pn spectra based on the three test models using the XSPEC {\tt fakeit} command. The simulated spectra were then optimally rebinned adopting the \citet{2016A&A...587A.151K} optimal binning scheme, just like the real data. Finally, we fit the DEM model to these simulated spectra (we fit the 0.35--2.5\,keV RGS and the 0.3--10\,keV EPIC-pn spectra simultaneously). Figure \ref{fig:test} a--c shows the fitting results. We found the DEM model can reproduce the one, two, and three temperature components very well, {as shown} by the single, double, and triple peaks seen in the fitting results. However, we noticed that the fitted peaks can be slightly shifted from the ``true values''. For example in Figure \ref{fig:test} c, the 4\,keV peak is shifted to $\sim5$\,keV, and the 1\,keV peak is also shifted a little bit to higher temperature. {A possible explanation on this could be that the 0.3\,keV and 1\,keV peaks are not perfectly separated by the DEM fitting, and thus the low-temperature EM may be slightly overestimated, which pushes the 1\,keV and 4\,keV peaks a little bit to higher temperatures. The degeneracy between temperature and ionization parameter may contribute as well. This level of systematic uncertainty will not alter the conclusions of this work.}

%\lsun{response to SO ``What concerns me (not for this work, but in general) is that we know the first bump is an artificial structure due to the simulation results. In scenarios where MHD simulations are not available, how can we identify these fake structures? Does this fake structure disappear if you exclude from the analysis the spectral regions with low S/N ratio?'':

%Your concern is necessary. One of the benefits of incorporating MHD simulations into this study is their ability to help verify the systematic uncertainty of the DEM model and identify any artificial features in the results. In fact, such anomalous features at very low temperatures have been already noticed very early in this study when I first applied the DEM model to SN 1987A, even prior to receiving the MHD simulation results from you. At that point, the origin of these features was unclear --- I suspected they were artificial structures because they consistently appeared at low temperatures, resembling some kind of boundary effect. However, after obtaining and testing with the synthetic spectra, I can be pretty sure that these features are artifacts. Therefore, MHD simulations are indeed very important for this work as a calibrator. On the other hand, it is also the specialty of SN 1987A that provides us with this opportunity to perform the calibration, which will be a useful reference for the DEM modelings of other SNRs with no MHD simulations available.}

The above test models are still quite simple and idealized, while the real X-ray gas in SN 1987A could have a much more complicated temperature structure. In order to {examine} the performance of the DEM model in a more complex situation, we adopted the synthetic {80\,ks} RGS (the 1st order spectrum of RGS1) and EPIC-pn spectra of SN 1987A at its age of {27} in 2014 from \citet{2022ApJ...931..132G} as another test. The spectra were synthesized based on the B18.3 model in \citet{2020A&A...636A..22O} and \citet{2020ApJ...888..111O}, which has been well constrained by multiwavelength observations of the progenitor star, the SN explosion, and the SNR. %%\lsun{[response to MM ``{\it - If I understand correctly,  the exposure time is 80 ks also in this case, is it correct?}'': Yes, you are right. I have added this information in text.]} 
The synthetic spectra represent a wide temperature distribution ranging in $\sim0.2$--$5$\,keV, as shown by the red line in Figure \ref{fig:test} d. The DEM fitting basically reproduces the temperature distribution with a main peak at $\sim0.6$\,keV and a high-temperature tail extending to $\sim4$\,keV. Some fine structures in the original distribution has been smoothed out due to the DEM model resolution limit. There is a fake structure seen at very low ({0.1--0.2\,keV}) temperature. Similar features can also be seen in some cases when we apply the DEM model to the real SN 1987A data (see Section \ref{sec:fit_real_data} for details). {Given the large uncertainties of the DEM at low temperatures,} we consider them as {most likely the} fake features caused by low signal-to-noise ratio {of the spectra} at very low energy band. {We note that given their rather low temperatures, these (fake) features make little contribution to the total X-ray emission, and thus will not alter the DEM profile at higher temperatures and the conclusions of this work.}
%%\lsun{[response to EG ``{\it - Did you try to fit the data with the C-stat instead of the classical $\chi^2$? This might help in some of the spectral regions with low $S/N$ ratio. On the other hand, I was wondering if you could also explain the fake bump between 1 and 2 keV, if it can be corrected somehow.}'': As I mentioned at the beginning of Section 3, I adopted the C-stat for all the spectral fittings in this work. I have no good explanation on the fake bump between 1 and 2 keV. However, it looks not as significant as the $<0.2$\,keV one.]}

With the tests above, we demonstrate the capability of DEM model in capturing the complex temperature structure of SN 1987A. No matter the X-ray gas is concentrated in two or three major components, or has a more continuous distribution, the DEM model is able to reproduce these features with good accuracy and reliability.

\subsection{{Application} to Real Data}\label{sec:fit_real_data}

\begin{figure*}[ht]
    \centering
    \includegraphics[width=\textwidth]{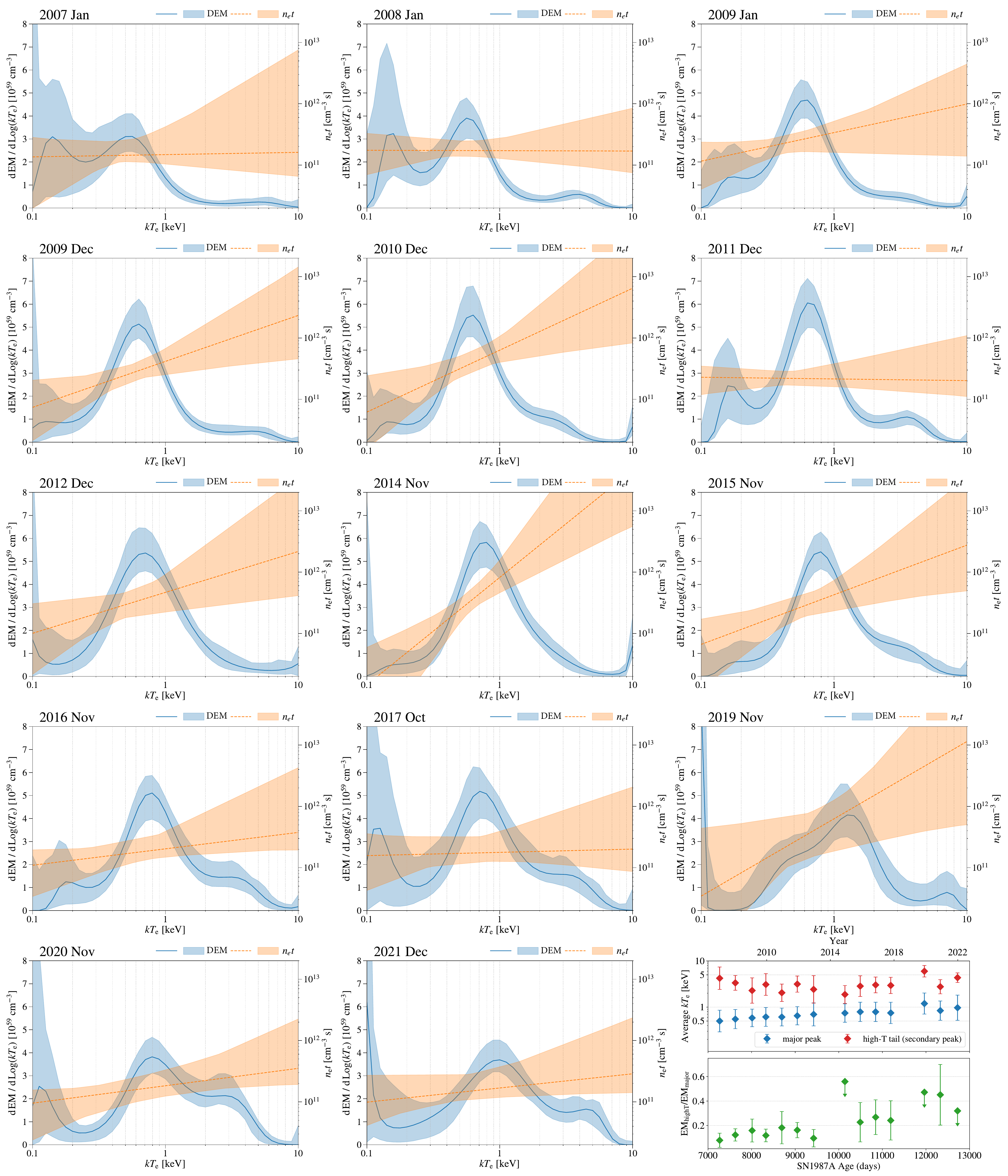}
    \caption{DEM fitting results for SN 1987A in different epochs based on XMM-Newton RGS and EPIC-pn observations. {In the bottom right panel we present the average temperatures of the major peak and the high-temperature tail (the blue and red data points in the upper sub-panel, respectively), and the ratio between the EMs of the two components (the green data points in the lower sub-panel).}}
    \label{fig:dem}
\end{figure*}

\begin{figure*}[ht]
    \centering
    \includegraphics[width=1\linewidth]{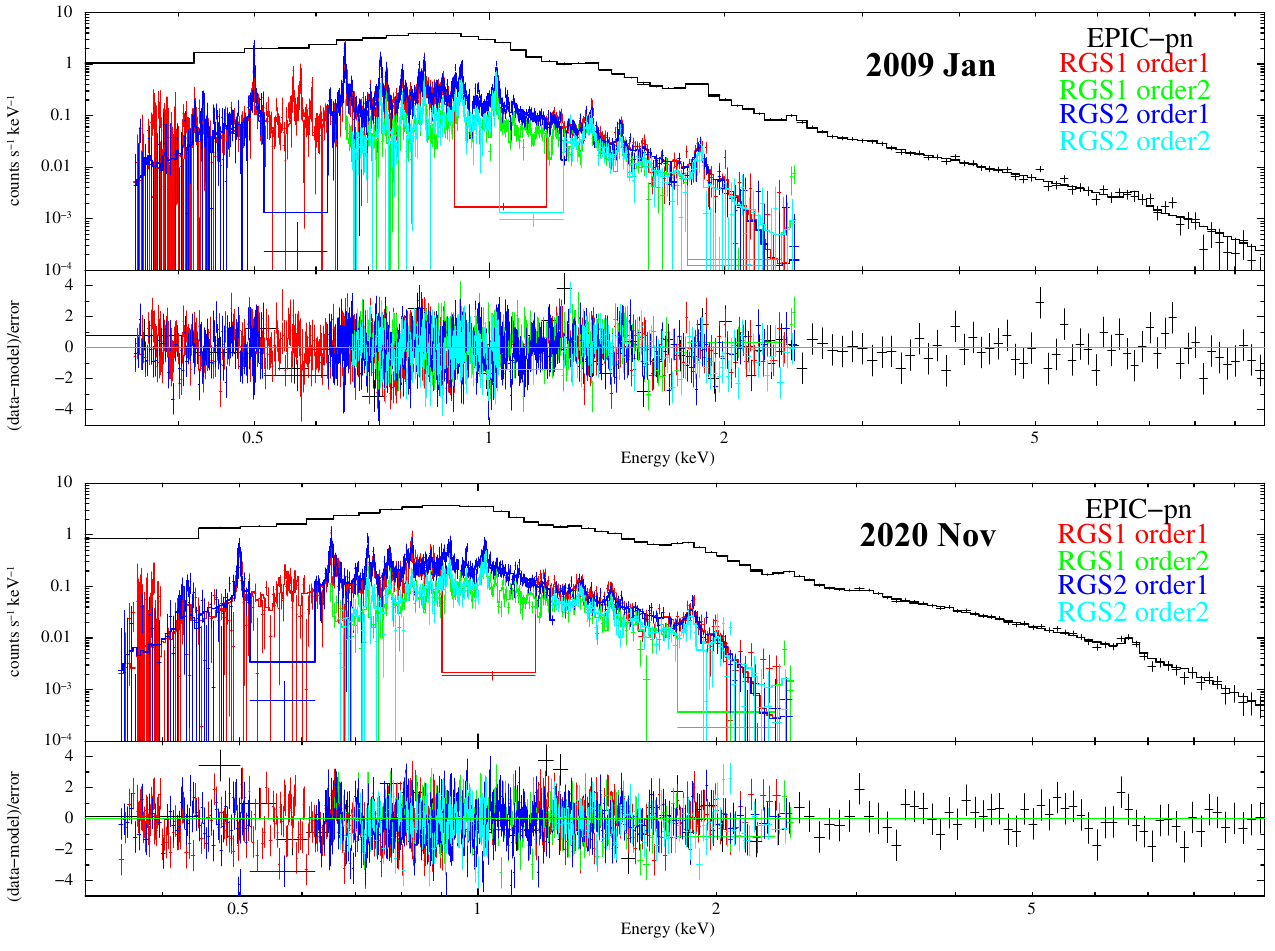}
    \caption{{Two examples of the fitted RGS + EPIC-pn spectra with residuals, taken from 2009 Jan and 2020 Nov. The spectra from other epochs are presented in Appendix \ref{app:corner_plot}.}}
    \label{fig:spec_example}
\end{figure*}

We applied our DEM model to each observation listed in Table \ref{tab:obs}, by fitting the 0.35--2.5\,keV RGS and the 0.3--10\,keV EPIC-pn spectra simultaneously. The obtained temperature distribution of the X-ray gas and its temporal evolution in 2007--2021 are shown in Figure \ref{fig:dem}. {The examples of the fitted spectra are shown in Figure \ref{fig:spec_example}.} We leave the detailed best-fit parameters, their 1-$\sigma$ uncertainties, the MCMC corner plots, {and the fitted spectra from all the epochs} in Appendix \ref{app:corner_plot}. The DEM model provides significantly better fit to the data than the discrete-temperature models. The C statistics ranges from 1796 to 2246 with d.o.f. ranging from 1584 to 1766. With only one additional parameter (thereby $\Delta{\rm d.o.f.}=-1$), the DEM model achieves an average improvement of the C statistic of $\overline{\Delta C}\sim-38$ (ranging from $-4$ to $-109$) compared with a 3-T {\tt vnei} model. {However, we note that despite significant improvements, the DEM modeling is still far from perfect and {leaves} some residuals in the fitted spectra. {One of the notable residual features is observed} at the O lines, which may {be} due to other mechanisms such {as} charge exchange, resonant scattering, and absorption of the foreground hot gas. We leave the discussions on O lines in {a companion} paper \citep{2025inprep}.} 

We found the X-ray gas in SN 1987A generally follows a similar temperature distribution during the $\sim15$ years. Rather than concentrating in two or three {well-separated peaks, it centers on only one major peak at $\sim0.5$--1\,keV in most of the epochs, while there is always a tail extending to the high-temperature end (up to $\gtrsim5$\,keV).}

{In order to better identify the major peak and the high-temperature tail, and estimate their mean temperatures and total EMs,  we fitted the obtained DEM distributions with two Gaussian profiles. We found that, for all the epochs, the double-Gaussian profile was capable of fitting the DEM distribution, and thus two temperature components (namely the major peak and the high-temperature tail) were identified. The average temperature of the two components, as well as the EM ratio between them (${\rm EM}_{\rm highT}/{\rm EM}_{\rm major}$), are shown in the bottom right panel of Figure \ref{fig:dem}. The average temperature of the major peak increased from $\sim0.5$\,keV to $\sim1$\,keV over a decade. The high-temperature tail has an average temperature $\sim2$--$5$\,keV, while no significant variation patterns can be observed.}
%The major peak gradually moved from $\sim0.5$\,keV to $\sim1$\,keV, indicating an increase in the average temperature of the plasma. 
The total EM of the major peak kept climbing in the first few years, reached its maximum at around 2011--2014, and then started to decline. This is consistent with previous results on the soft band (e.g., 0.5--2\,keV) light curves \citep[e.g.,][]{2016ApJ...829...40F,2021ApJ...916...41S,2021ApJ...916...76A,2024ApJ...966..147R} and the low-temperature plasma component EMs \citep[e.g.,][]{2021ApJ...916...41S,2021ApJ...916...76A,2021ApJ...922..140R}. 
On the other hand, the high-temperature tail kept existing and increasing in its EM during the whole period.
{As a result, the EM ratio between the high-temperature tail and the major peak has been continuously increasing, from $\sim0.1$ at around 2007 to $\gtrsim0.2$ after 2014, and reaching $\sim0.4$ at around 2020 (the bottom right panel in Figure \ref{fig:dem}). This suggests a greater contribution of the high-temperature plasma to the X-ray emission in the recent years. As a consequence, the high-temperature tail becomes more and more significant in the DEM profile, and appears to have formed a secondary peak in 2020--2021.}

In the DEM distribution of some observations (e.g., 2008, 2011, and 2017), we found {a} potential peak at very low temperature ($\lesssim0.2$\,keV) with large uncertainties. As we have discussed in Section \ref{sec:test}, these are most likely the fake features caused by low signal-to-noise ratio of the spectra at very low energy band. However, it is still possible that these features are from a real plasma component with a rather low temperature and high density (therefore a low shock velocity to achieve a low post-shock temperature). We note that such a component has been observed at the early stage of the X-ray remnant of SN 1987A, when the SN shock just encountered with the dense {inner} part of the equatorial ring. For example, \citet{2004ApJ...610..275P} reported a plasma component with $kT_e\sim0.22$\,keV and $n_e\sim7500$\,cm$^{-3}$ resulted by the slow transmitted shock into the dense inner ring. {Another possible interpretation of this low-temperature component could be the shocked stellar wind from the BSG progenitor of SN 1987A. The typical temperature of such kind of shocked stellar wind is $\sim0.1$\,keV \citep[e.g.,][]{1975ApJ...200L.107C}. At last, the emission from the local hot bubble, if has not been fully subtracted during the background subtraction process, may also contributes to this low-temperature component. {Nevertheless, the fact that this low-temperature feature comes and goes with time is still against the interpretation as a real shock component.}}

\section{Discussions}\label{sec:discussion}

\subsection{{Comparison with MHD simulations: Interpretation of the DEM results}}\label{sec:obs_vs_simulation}

\begin{figure*}[ht]
    \centering
    \includegraphics[width=\textwidth]{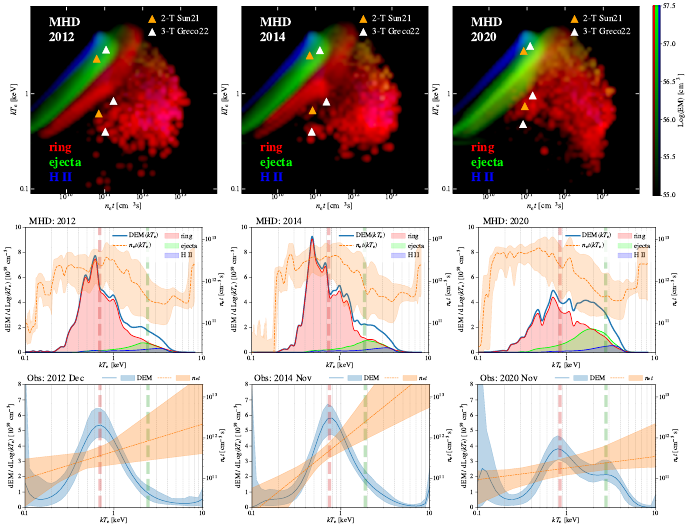}
    \caption{{Comparison between MHD simulations and the DEM fitting results of SN 1987A. {\it Top}: simulated plasma EM distributions in $kT_e$--$n_et$ diagrams \citep[based on the B18.3 model in][]{2020A&A...636A..22O}, overlaid with 2-T \citep{2021ApJ...916...41S} and 3-T \citep{2022ApJ...931..132G} spectral fitting results. Different components (i.e., ring, ejecta, and H II region) are indicated by different colors (i.e., red, green, and blue). {\it Middle}: simulated DEM distributions (blue solid lines) and EM-weighted $n_et$ distributions (orange dashed lines). {\it Bottom}: DEM fitting results, same as those in Figure \ref{fig:dem}. {The red and green dashed lines in the middle and bottom panels denote the average temperatures of the major peak and the high-temperature tail (the secondary peak) obtained from the observations, respectively.}}
%    \resolved{The first and second row can be immediately improved by involving contributions from different components (i.e., the H II region, ER, and ejecta). Is that possible for Emanuele or Salvatore to kindly provide the EM from different components (just like those in Greco+2022), if you still have it?A question for Emanuele and Salvatore: I'm not sure whether the EM distributions (shown in the first row) have been smoothed or not, and if yes, whether the smoothing keeps the total EM unchanged thus will not affect the $kT_{\rm e}$ \& $n_{\rm e}t$ distributions (shown in the second row)?}
    }
    \label{fig:MHD_obs}
\end{figure*}

SN 1987A is one of the few SNRs for which people have performed comprehensive 3D {MHD} simulations tracing their long-term evolution from the onset of the SN all the way to the remnant phase \citep[e.g.,][]{2020A&A...636A..22O}. It provides us with an unprecedented opportunity to connect simulations with observations --- not only in the sense that the long-term, multiwavelength observations of SN 1987A (including its progenitor star, the SN explosion, and its remnant) set detailed constraints on the initial conditions and help to regulate the simulations; but also in the sense that the simulation results, together with the synthesized observables (including the multiwavelength images, light curves, and spectra), are well comparable with observations, which may in turn help us to better understand the observations and make predictions.

One of the most impressive outcomes of the SN 1987A simulations is that {it shows} how well the synthesized X-ray spectra match with the actual observations. \citet{2022ApJ...931..132G} synthesized the SN 1987A X-ray spectra in three epochs (2012, 2014, and 2020) based on the B18.3 model in \citet{2020A&A...636A..22O} and \citet{2020ApJ...888..111O}, and compared them with XMM-Newton RGS, XMM-Newton EPIC-pn, Chandra ACIS-S, and NuSTAR FPMA observations. Without performing any fit, they found the synthesized thermal spectra reproduce the observed $<10$\,keV spectra very well at all epochs for all instruments \citep[see Figure 2 in][]{2022ApJ...931..132G}. 
The simulations provide a continuous distribution of the plasma in a large parameter space, which was adopted in X-ray spectra synthesis. The first row of Figure \ref{fig:MHD_obs} shows the simulated plasma EM distributions of SN 1987A in the $kT_e$--$n_et$ diagram at different epochs. Spectral fitting results from previous studies adopting two-temperature \citep{2021ApJ...916...41S} and three-temperature \citep{2022ApJ...931..132G} models are also plotted in order to compare with simulations. We found that even though the 2-T and 3-T models may represent some kinds of average properties of the plasma, they are still too simplified to describe the complex structure of the remnant.

Our DEM modeling results provide a unique perspective to compare observations with simulations. The second and third rows of Figure \ref{fig:MHD_obs} present the DEM distributions of the X-ray gas in SN 1987A obtained from simulations and observations, respectively. At 2012 and 2014, simulations indicate that the majority of EM concentrates at $\sim0.6$\,keV, which is dominated by the ER and forms a major peak in the DEM curve. The shocked H II region and ejecta materials have higher temperatures (up to $\sim5$\,keV), which result in a tail in the DEM distribution to the high-temperature ends. In 2020, simulations predict a decrease in the EM of the ER-dominated low-temperature plasmas, resulting in a weakening of the major peak. On the other hand, the amount of the shock-heated ejecta materials has been significantly increasing, which raises the high-temperature tail and makes it comparable with the major peak. The DEM fitting results agree well with the simulations in almost all respects: the temperature and the EM of the major peak, the existence of a high-temperature tail, and the overall evolution of the DEM profile. This consistency not only strengthens the reliability of both sides, but also {helps} us to get a better understanding of SN 1987A. In particular, as described above, we found a dramatic decline of the major peak and a rapid rising of the tail (the emergence of a secondary high-temperature peak) in the recent few years based on DEM analysis. Together with simulation results, this reveals the fading of the ER and the brightening of the shocked ejecta.

{Despite the general consistency in most aspects, there are still some discrepancies between the MHD simulations and the observations, such like the average temperatures of the major peak and the high-temperature tail (the secondary peak), the absolute value of the DEM, and the $n_{\rm e}t$ distributions. The observed average temperatures of the major and secondary peak appear to be slightly higher than those in simulations, which may suggest a higher shock velocity. The absolute DEM derived from our modeling is slightly lower than the simulation expectations. This may indicate an overall lower density of the shocked gas. However, we note that the simplified DEM model may smooth out the fine structures, leading to lower average values. Additionally, the difference in metal abundances may also lead to different DEM values. Our DEM modeling suggests a flat or positive correlation between $n_{\rm e}t$ and $kT_{\rm e}$ (i.e., $\beta\gtrsim0$) for most of the observations. The EM-weighted $n_{\rm e}t$ according to simulations exhibits an increase with $kT_{\rm e}$ at very low temperatures ($\lesssim0.3$\,keV), a plateau around $\sim0.3$--$1$\,keV, then a decrease in $\sim1$--$5$\,keV, and finally an increase again at very high temperatures ($\gtrsim5$\,keV). However, the uncertainty of the fitted $n_{\rm e}t$ and the scattering of the simulated $n_{\rm e}t$ are both large. We leave a detailed discussion on the $n_{\rm e}t$--$kT_{\rm e}$ correlation in Section \ref{sec:n_et}.}

\begin{figure*}
    \centering
    \includegraphics[width=1\linewidth]{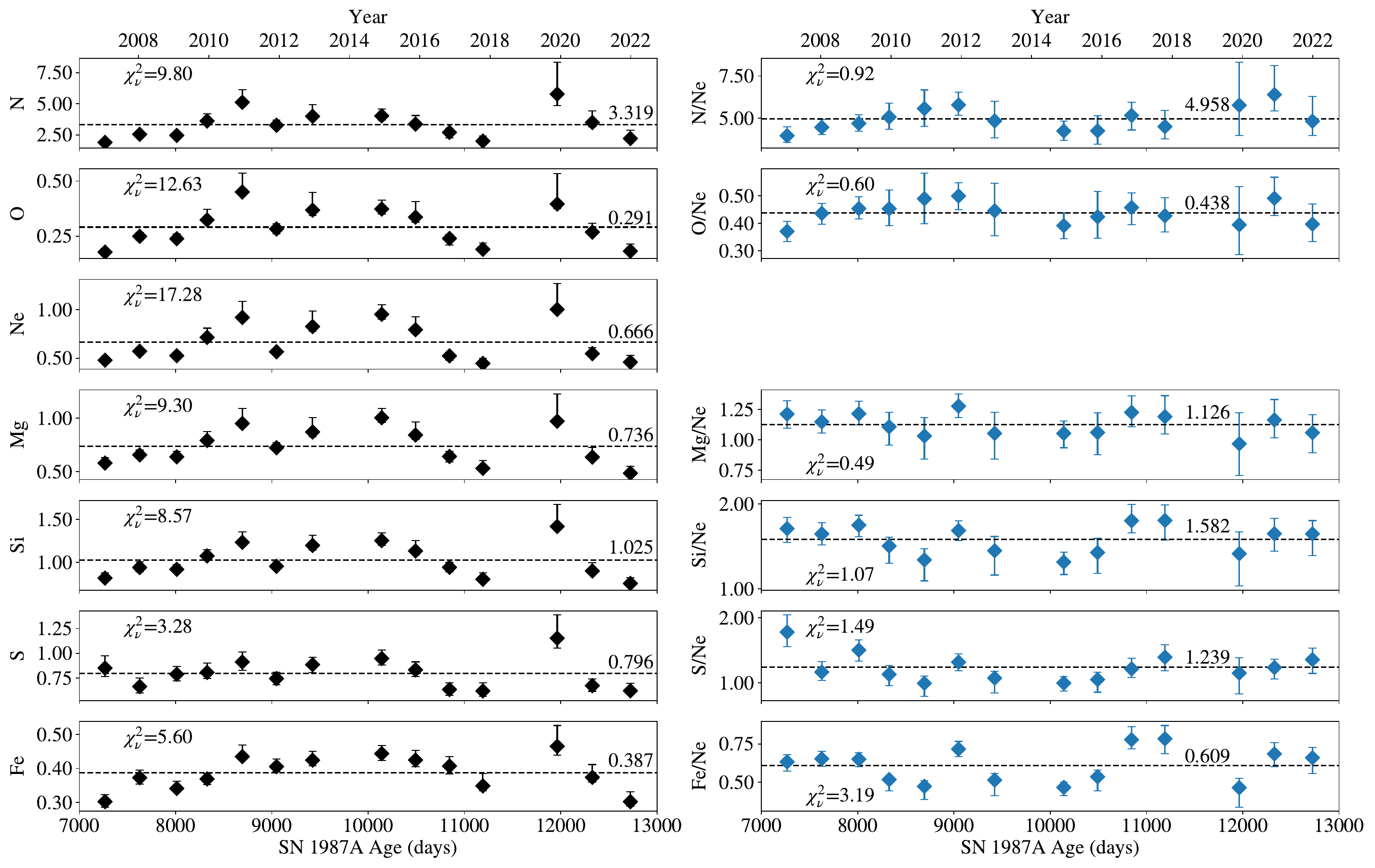}
    \caption{Metal abundances (left) and abundance ratios with respect to Ne (right) in SN 1987A. The black dashed lines indicate the mean values. Reduced chi-squares $\chi_{\rm \nu}^2=\chi^2/{\rm dof}$ are calculated based on the mean values. For a $p$-value of 0.002 ($\sim3\sigma$ level), the critical reduced chi-square is $\chi^2_{\rm \nu,c}\approx2.50$ for 13 degrees of freedom.}
    \label{fig:abundances}
\end{figure*}

Recently, \citet{2022A&A...661A..30M} performed an in-depth analysis on the XMM-Newton EPIC-pn observations of SN 1987A, by simultaneously fitting the spectra from all epochs with a three-temperature plasma model. They considered different metal abundances between different components, and found that the abundances of the ``hot'' component are significantly higher than those of the ``warm'' and ``cooler'' components, for which they suggested the ``hot'' component to be originated from the reverse shock heated ejecta. Figure \ref{fig:abundances} shows the metal abundances obtained from our DEM analysis. In our DEM model, the abundances are assumed to be identical at different temperatures, in order to limit the number of free parameters and to keep the model as simple as possible. Therefore the abundances we obtained should be taken as the (EM-weighted) average values in SN 1987A. Even though the MHD simulations and the DEM distributions both indicate a recently brightened ejecta component, we didn't {find} an increase in the average metal abundances. On the contrary, after a plateau around 2009--2016 ($\sim8000$--11000\,days), the abundances have been gradually decreasing and recently got back to the level as they were around 2007 ($\sim7000$\,days). This is consistent with the findings by \citet{2021ApJ...916...41S} based on 2-T fitting, and indicates that the ER has a higher metal abundances (or say a lower hydrogen context) compared with the ambient H II region, and is now fading out. Although the shocked ejecta becomes increasingly dominant in X-ray emission, {the decreasing average metal abundances suggest that} {the} reverse shock has not yet reach the inner metal-rich core. This is also supported by simulations and optical observations, which show the outermost layer of the ejecta is dominated by hydrogen-rich materials \citep[e.g.,][]{2013ApJ...768...88F,2016ApJ...833..147L,2020ApJ...888..111O,2020A&A...636A..22O}. {However, the most recent JWST NIRSpec observations indicate that the inner Fe-rich ejecta are now starting to interact with {the} reverse shock \citep{2023ApJ...949L..27L}. Therefore we are expecting to see the corresponding X-ray emission in the near future.}

The metal abundances we obtained with DEM modeling are systematically higher than those obtained by \citet{2021ApJ...916...41S} and \citet{2022ApJ...931..132G} based on 2-T and 3-T fittings. {Furthermore} we note that the 3-T models tend to give slightly higher abundances than 2-T models. This phenomenon has been noticed for a long time in X-ray spectroscopic studies, especially in those associated with the spectral fitting {for} ellipticals, groups and clusters of galaxies, which is referred as ``Fe bias'' or ``Si bias'' \citep[e.g.,][]{2000MNRAS.311..176B}. {It is} found that, for a plasma that has an intrinsic wide continuous temperature distribution, an over-simplified 1-T modeling of the spectrum may result in underestimated metal abundances. Introducing more temperature components will improve the fit and provide unbiased constraints on abundances. The underlying mechanism could be that, the 1-T models (or any discrete-temperature models) may overestimate the continuum emission, since part of the ``continuum'' is actually the pseudo-continuum formed by numerous faint emission lines from other (unconsidered) temperature components. {Similar phenomena has also been noticed and discussed in the context of pure-metal plasma by \citet{2020A&A...638A.101G}, where they suggested a high-resolution spectroscopic study, especially on the resolved radiative recombination continuum (RRC) emission, may help to disentangle the degeneracy between metal abundances and EM.} In this sense, our results provide an up-to-date most unbiased constraint on the average metal abundances of SN 1987A. %%\lsun{[response to EG ``{\it - In a paper that I published with Jacco, Marco and others in 2020 we showed that there is a degeneracy between emission measure (i.e. continuum emission) and absolute abundances that is intrinsic to the spectra. It might be worth it to mention this somewhere in this paragraph, if you think it can be useful.}'': Thank you for mentioning that! The abundance-EM degeneracy discussed in your work and the discussions here share the same underlying mechanism. We should certainly mention your work here.]}

\subsection{Comparison with previous studies: improvements and advantages} \label{sec:compare}

Most of the previous studies on SN 1987A adopted discrete-temperature (either two-temperature or three-temperature) plasma models to fit its X-ray spectra \citep[e.g.,][]{2006A&A...460..811H,2004ApJ...610..275P,2006ApJ...646.1001P,2006ApJ...645..293Z,2009ApJ...692.1190Z,2010MNRAS.407.1157Z,2008ApJ...676..361H,2010A&A...515A...5S,2012ApJ...752..103D,2016ApJ...829...40F,2020ApJ...899...21B,2021ApJ...916...76A,2021ApJ...916...41S,2021ApJ...922..140R,2024ApJ...966..147R,2021ApJ...908L..45G,2022ApJ...931..132G,2022A&A...661A..30M}. As summarized by \citet[][see Table 6 therein]{2021ApJ...916...76A}, these works found plasma temperatures mostly in the range of $\sim0.5$--$4$\,keV, with a low-temperature component $\sim0.5$\,keV, a high-temperature component $\sim2$--$4$\,keV, and some of them an intermediate component $\sim1$\,keV. As mentioned in Section \ref{sec:obs_vs_simulation}, even though these discrete-temperature models may in some way characterize the average properties of the plasma, they are still over-simplified, and we are still looking for a better approach to describe the complex structure of the remnant and to compare with simulations.

For now, only a few attempts have been made to model the continuous temperature distribution of SN 1987A. \citet{2006ApJ...645..293Z,2009ApJ...692.1190Z} made the first try by applying a ``distribution of shocks (DS)'' model (which is very similar to our DEM model described in Section \ref{sec:DEM_model}, but is based on {\tt vpshock} rather than {\tt vnei}) to Chandra LETG/HETG data taken in 2004 and 2007. They found a bimodal temperature distribution that peaks at $\sim0.5$ and $\sim2.2$--$3$\,keV. Recently, \citet{2021ApJ...916...76A} performed a similar analysis on RGS $+$ NuSTAR data taken in 2012 to 2020. However, their results suggested three peaks in the continuous temperature distribution, at $\sim0.3$, $\sim0.9$, and $\sim4$\,keV, respectively (the separation between the $0.3$ and $0.9$\,keV peaks is marginally significant). The results from above studies are not consistent with each other (which could be due to the observations they used are from different epochs and instruments), and neither of them agree well with the simulations.

Our results {are different} from \citet{2006ApJ...645..293Z,2009ApJ...692.1190Z} and \citet{2021ApJ...916...76A}, while some similarities can be found. Rather than well-separated two or three peaks, we found only one major peak in most of the observations. The major peak is quite wide, overlapping with the $0.5$\,keV peak in \cite{2006ApJ...645..293Z,2009ApJ...692.1190Z} and the $0.3$ and $0.9$\,keV peaks in \citet{2021ApJ...916...76A}. The high-temperature peak is not very significant in our results until 2020, which appears as a tail extending to $\sim5$\,keV in most of the observations. For a detailed comparison with previous studies, we summarize the {major} improvements and advantages of our work as below:
\begin{itemize}
    \item {\bf Data}
    
    \citet{2006ApJ...645..293Z,2009ApJ...692.1190Z} adopted Chandra LETG/HETG observations of 2004 and 2007. Limited by the relatively low effective area and the low luminosity of SN 1987A at that time, the spectra may have lower statistics compared with those taken by XMM-Newton RGS and EPIC-pn after 2007. In particular, they may have got less constraints on the hard band continuum, and thus the high-temperature plasma components. \citet{2021ApJ...916...76A} jointly fitted both the RGS and NuSTAR spectra, which provide a much wider energy coverage and result in good constraints on both the low- and high-temperature gas. However, since the spectra in different energy bands are from different observations taken by different instruments, their results may subject to systematic uncertainties arising from, e.g., temporal variation of SN 1987A between RGS and NuSTAR observations taken in different epochs\footnote{The time differences between XMM-Newton and NuSTAR observations adopted by \citet{2021ApJ...916...76A} for their joint fit are generally around $\sim100$--$200$\,days, whereas previous studies have showed that the X-ray flux (either broad-band flux or line flux) and the surface brightness distribution of SN 1987A varies in a timescale of a few hundred days \citep[e.g.,][]{2016ApJ...829...40F,2021ApJ...916...41S}.}, cross-calibration uncertainties between RGS and NuSTAR, and the energy gap between RGS and NuSTAR spectra (at $\sim2$--$3$\,keV, which leads to a missing of the S lines and the underlying continuum). {We note that some of these systematic uncertainties have been investigated in Appendix D of \citet{2021ApJ...916...76A}. They simultaneously fitted the RGS + EPIC-pn + NuSTAR data from several epoches with their continuous-temperature model, and introduced a variable cross-normalization constant for each instrument in order to account for cross-calibration uncertainties. They found that including the EPIC-pn data does not qualitatively affect their conclusions, though there could be a hint that the two low-temperature peaks are less distinct \citep[see the upper left panel of their figure D2 in][]{2021ApJ...916...76A}.}
    In this work, we jointly fitted the RGS $+$ EPIC-pn spectra, which are taken at the same time for each observation and cover a wide energy range from 0.35\,keV to 10\,keV {without any} gap. Our DEM analysis characterizes the continuous temperature distribution of the hot gas and its temporal variation from 2007 to 2021, thus provides up to now the most complete picture on the X-ray evolution of SN 1987A since SN shock encountered with the main ER.
    {We confined our analysis to XMM-Newton observations while did not adopted the NuSTAR data, in pursuit of minimizing the systematic uncertainties introduced by different observations from different instruments. However, this leads to a disregarding of the $>10$\,keV spectra. We note that, despite the controversy, based on a simultaneously analysis of the Chandra, XMM-Newton, and NuSTAR data, \citet{2021ApJ...908L..45G,2022ApJ...931..132G} suggested that at $>10$\,keV, the X-ray emission of SN 1987A is dominated by the non-thermal emission from the central pulsar wind nebula (PWN). Due to the large column density ($N_{\rm H}>10^{23}$\,cm$^{-2}$) of the unshocked ejecta, non-thermal emission from the PWN subjected to heavy absorption, and thus has no significant effect to the $<10$\,keV spectra which is dominated by thermal emission from the shocked hot plasma. Given the fact that our DEM analysis was aimed at the thermal-emitting gas, we expect that disregarding the NuSTAR data will not significantly affect our major conclusions. On the other hand, it is possible that the $>10$\,keV emission is contributed by a high-temperature plasma component, i.e., $\sim4$\,keV as suggested by \citet{2021ApJ...916...76A}. However, we note that such high-temperature plasma component has already been characterized in our DEM modeling, even without adopting the NuSTAR data. Nevertheless, a conjoint analysis of RGS + EPIC/Chandra + NuSTAR data may help to set tighter constraints on both the thermal-emitting gas and the non-thermal-emitting PWN, which is of great interest but is out of the scope of this paper.}
    
    \item {\bf {Spectral model}}
    
    The spectral models used in \citet{2006ApJ...645..293Z,2009ApJ...692.1190Z}, \citet{2021ApJ...916...76A}, and our study are quite similar. The major difference could be that our DEM model is based on the {\tt vnei} model in XSPEC, while previous ones are based on {\tt vpshock}. The {\tt vpshock} model assumes a plane-parallel shock, where the post-shock gas has a linear distribution of the ionization parameter $\tau$ versus the EM between a lower limit $\tau_{\rm l}$ and a upper limit $\tau_{\rm u}$ \citep{2001ApJ...548..820B}. 
    {On the other hand, the {\tt vnei} model characterizes the average property of the plasma with a single temperature $kT_{\rm e}$ and a single ionization parameter $\tau$. Therefore, {\tt vpshock} model provides a more accurate (but still highly simplified and idealized) description for the plasma whose actual physical condition is (expected to be) close to that under the plane-parallel shock assumption. However, for a complex system such like SN 1987A, the integrated plasma property may significantly deviate from the plane-parallel shock schema with a much more complicated thermalization and ionization history (e.g., as shown by MHD simulations in the first row of Figure \ref{fig:MHD_obs}). In this case, the {\tt vnei} model, although even more simplified compared with {\tt vpshock}, may be more flexible and viable in the sense of providing a general characterization of the average property of the plasma.} We have also constructed a DEM model based on {\tt vpshock}, and found that the obtained C statistic and best-fit parameters are similar to those using {\tt vnei}-based DEM model ({details on the {\tt vpshock}-based DEM modeling are given in Appendix \ref{apx:vpshock}}). \citet{2006ApJ...645..293Z,2009ApJ...692.1190Z} restrict the plasma temperature to $0.15$--$4$\,keV, which may limit the constraint on high-temperature components. Both \citet{2006ApJ...645..293Z,2009ApJ...692.1190Z} and \citet{2021ApJ...916...76A} adopted a maximum order Chebyshev series $M=6$ and a total temperature bin number $N=25$, which in our case are $M=7$ and $N=40$ {instead}. As mentioned in Section \ref{sec:DEM_model} {and detailed in Appendix \ref{app:M_N}}, this setup has been optimized for our data through multiple tests, therefore may provide better performance. At last, \citet{2006ApJ...645..293Z,2009ApJ...692.1190Z} was based on a rather old version of atomic database ({the {\tt nei} model version 2.0 in XSPEC version 11.3.2, which is based on AtomDB version 2.0}), which is less accurate and may subject to some errors.
    
    \item {\bf Procedure}
    
    Neither \citet{2006ApJ...645..293Z,2009ApJ...692.1190Z} nor \citet{2021ApJ...916...76A} presented the verification procedure of the model, and they didn't provide the uncertainties ({i.e.,} error bars) of the obtained temperature distributions. We first verified the validity and the reliability of the DEM model through multiple tests as in Section \ref{sec:test}. {In particular, the MHD simulations provide us with unique opportunity to verify the systematic uncertainty of the DEM model and to identify any artificial features in the results.} Then with an MCMC approach, we {obtained} not only the best-fit curves but also the 1-$\sigma$ uncertainties of the temperature distributions.
\end{itemize}

{In addition to the major improvements summarized above, there could be other small differences lying in the detailed data processing and analyzing procedures which may not be fully enumerated. Among all the differences, it is likely that the the differences between the data set that being used play the most important role in causing the discrepancies in the final results, given that \citet{2006ApJ...645..293Z,2009ApJ...692.1190Z} and \citet{2021ApJ...916...76A} obtained different DEM distributions based on different data set even if they adopted quite similar model configurations.}

\subsection{{Correlation between $n_{\rm e}t$ and $kT_{\rm e}$}}\label{sec:n_et}

%%\lsun{This section should be improved after discussing with collaborators.}

% adiabatic cooling/heating
% post-shock electron-ion temperature equilibrium
% temperature-density relation

The hot plasma in SN 1987A is under the NEI state, i.e., the DEM is not only a function of electron temperature $kT_{\rm e}$, but also a function of ionization parameter $n_{\rm e}t$. Therefore, one of the major challenges in our DEM modeling is how to deal with $n_{\rm e}t$. In order to get a better constraint on the temperature distribution (which is the major objective of this work) and to keep the model as simple as possible, we made the compromise that assumed a simple power-law relation between $n_{\rm e}t$ and $kT_{\rm e}$ {(see Eq. \ref{eq:net})}. However, this assumption, which has also been adopted by \citet{2006ApJ...645..293Z,2009ApJ...692.1190Z} and \citet{2021ApJ...916...76A}, might be questionable, given the rather complicated thermal and ionization structure of the plasma.

In the case of shock transmitting into a density gradient, we have $kT_{\rm e}\propto V_{\rm sh}^2\propto n_{\rm e}^{-1}$, where $V_{\rm sh}$ is the shock velocity. If we assume the plasmas with different densities are shocked at the same time, then there will be a simple relation between temperature and ionization parameter as $kT_{\rm e}\propto (n_{\rm e}t)^{-1}$ (i.e., $\beta=-1$ in Eq. \ref{eq:net}). 

In collisionless shock heating, the post-shock temperatures of different particle species can be different, and the electron-ion temperature ratio can be significantly lower than one, especially in the case of high-Mach-number shocks \citep[e.g.,][]{2007ApJ...654L..69G,2023ApJ...949...50R}. Observational evidences of this temperature non-equilibrium effect have been found for several SNRs, and especially for SN 1987A \citep{2019NatAs...3..236M}. The post-shock electron temperature will then be gradually increased as interacting with other species, and finally get to equilibrium. The increasing rate $dkT_{\rm e}/dt$, taking electron-proton equilibration as an example, is approximately proportional to proton density $n_{\rm p}$ \citep[e.g.,][]{2020pesr.book.....V}. Given $n_{\rm p}\approx n_{\rm e}$, in this sense we got $kT_{\rm e}\propto n_{\rm e}t$ (i.e., $\beta=1$).

There are still many other mechanisms which may modify the $kT_{\rm e}$-$n_{\rm e}t$ relation, such as adiabatic cooling, thermal conduction, radiative cooling, etc. {Furthermore, as revealed by MHD simulations (Figure \ref{fig:MHD_obs}), at different scales and evolution stages, the plasma of different components could be dominated by different mechanisms, making the overall $kT_{\rm e}$-$n_{\rm e}t$ relation extremely complicated. According to simulations, the EM-weighted $n_{\rm e}t$ (Figure \ref{fig:MHD_obs} middle panels) shows an increase with $kT_{\rm e}$ at very low temperatures ($\lesssim0.3$\,keV), a plateau around $\sim0.3$--1\,keV, then a decrease in $\sim1$--$5$\,keV, and finally an increase again at very high temperatures ($\gtrsim5$\,keV). In addition to the complex relation between $n_{\rm e}t$ and $kT_{\rm e}$, the intrinsic scattering of $n_{\rm e}t$ is quite large.} Therefore, the power-law assumption adopted in the DEM modeling {might} be over-simplified, and the results must be treated with caution.

Our DEM modeling suggests a flat or positive correlation between $n_{\rm e}t$ and $kT_{\rm e}$ (i.e., $\beta\gtrsim0$) for most of the observations (Figure \ref{fig:dem}). {However, the uncertainties of $n_{\rm e}t$ are rather large, especially at the lower and higher ends of temperatures. The large uncertainties may be partially due to the poor statistics of the spectra at low and high energy bands. More importantly, it may be caused by the intrinsic scattering of $n_{\rm e}t$ at each temperature bin, and thus implies the complexity of the ionization history of the plasma. In our DEM modeling, $n_{\rm e}t$ is constrained mainly based on emission lines --- especially those lying in the $\sim0.5$--$2$\,keV band that are observed and well resolved by RGS. These emission lines are dominated by the plasma within the major peak ($kT_{\rm e}\sim0.5$--$1$\,keV). Therefore, despite the large uncertainties at the lower and higher ends of temperatures, the $n_{\rm e}t$ around $\sim0.5$--$1$\,keV may still be able to characterized the bulk ionization state of SN 1987A, which is dominated by the shocked ring.} On the other hand, the $n_{\rm e}t$ at 1\,keV has been decreasing in the recent few years, which may be associated with the newly shocked ejecta that lowers the average ionization state of the plasma.

\subsection{Latest evolution of Fe K line and its implications to newly shocked ejecta materials}

\begin{figure}[ht]
    \centering
    \includegraphics[width=\linewidth]{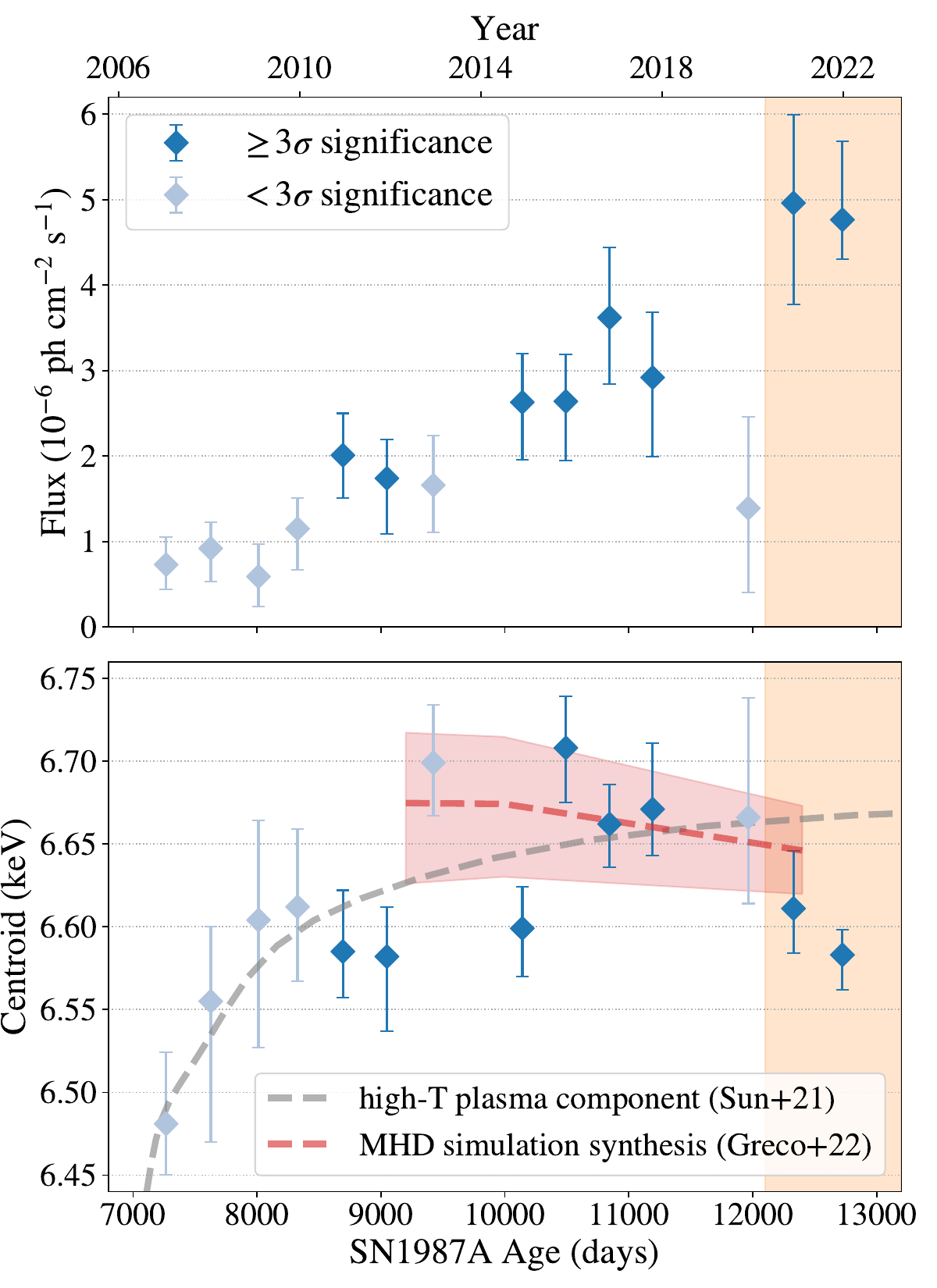}
    \caption{Light curve (upper panel) and centroid energy evolution (lower panel) of the Fe K line from SN 1987A. Data points before 2019 are adopted from \citet{2021ApJ...916...41S}, while the last two are new results (highlighted with orange areas). Points with $\gtrsim3\sigma$ detection are shown in dark blue, while the others in light blue. The gray dashed curve is adopted from \citet{2021ApJ...916...41S}, which represents the Fe K centroid energy for a high-temperature plasma component with $kT_{\rm e}=3.2$\,keV and $n_{\rm e}=500$\,cm$^{-3}$, shocked at 7000\,days after the explosion. The red dashed curve indicates the MHD simulation predicted centroid energy, obtained by fitting the synthetic spectra evaluated by \citet{2022ApJ...931..132G}.}
    \label{fig:Fe_K}
\end{figure}

{The newly shocked ejecta that are uncovered by our DEM analysis as well as the MHD simulation can also be supported by the recent changes of the Fe K line centroid.}
The Fe K complex contains a series of Fe K-shell fluorescent lines, where the line energy rises slowly in the range of $\sim6.4$--$6.7$\,keV as a function of ionization state \citep[e.g., Figure 12 in][]{2012A&ARv..20...49V}. The emission from individual ion species can hardly be resolved, and the whole Fe K complex usually appears as one single emission line in the CCD spectrum. The centroid energy of this Fe K line is thereby determined by the contributions of different ions, and thus may represent the average temperature and ionization parameter of Fe \citep[e.g., Figure 10 in][]{2021ApJ...916...41S}. 

Fe K line has been detected from SN 1987A using different instruments \citep[e.g.,][]{2010A&A...515A...5S,2012A&A...548L...3M,2021ApJ...916...76A,2021ApJ...916...41S,2022A&A...661A..30M,2024ApJ...966..147R}. In particular, \citet{2021ApJ...916...41S} found both of the flux and the centroid energy of Fe K line have been increasing from 2007 to 2019. Here, we update the latest variations of Fe K line based on new XMM-Newton EPIC-pn observations taken in 2020 and 2021. As shown in Figure \ref{fig:Fe_K}, the flux of Fe K line was still increasing, corresponding to the increasing EM of the high-temperature plasma. On the other hand, {we found a moderate decline of }the centroid energy from $\sim6.65$--$6.7$\,keV to $\lesssim6.60$\,keV in the last few years {(see Appendix \ref{apx:Fe_K} for a detailed investigation on the statistical significance)}, which {implies} the average ionization state of Fe {might have} been decreasing recently. As shown by \citet{2021ApJ...916...41S}, the temporal evolution of the Fe K centroid energy before 2019 can be well described by a plasma component with a temperature $kT_{\rm e}=3.2$\,keV and a density $n_{\rm e}=500$\,cm$^{-3}$, shocked at $\sim7000$\,days after the explosion, seen as the gray dashed curve in Figure \ref{fig:Fe_K}. {The recent drop-off in ionization state deviates from this scenario and indicates that a newly shocked plasma component may have contributed to the Fe K emission.}

\citet{2022ApJ...931..132G} presented a detailed discussion on the origins of Fe K line (see Section 3 and Figure 4 therein). Contributions from different components (i.e., the ER, the ejecta, and the H II region) at different epochs have been evaluated based on the MHD simulations from \citet{2020A&A...636A..22O}. The simulations suggest that the Fe K emission is firstly dominated by the shocked ring, especially by the lower-density material lying between the dense clumps, at around 2012. Then the contribution from the shocked ejecta gradually increases, which becomes comparable with the ring at around 2020 and will finally dominate the Fe K emission in the near future. Since the newly shocked ejecta has a lower average ionization parameter than the ring, this will result in a decline of the Fe K line centroid energy. This is demonstrated by the red dashed line in Figure \ref{fig:Fe_K}, which is obtained by fitting the synthetic spectra evaluated by \citet{2022ApJ...931..132G}. We found the MHD simulation synthesis that involves the complex constitution of the Fe-K-emitting gas, especially the recently shocked ejecta, can be better compared to observations.

In summary, {we found a moderate decrease of the Fe K centroid energy in the last few years which} can be well explained by simulations, and thus {may} provide another observational evidence for the brightening of the shocked ejecta, in addition to the DEM results.

{\citet{2024ApJ...966..147R} measured the fluxes and centroid energies of Fe K line from 2018 to 2022 using Chandra ACIS observations (see their Table 2 and Figure 6). They found the Fe K flux has significantly increased since 2018, from $\sim4\times10^{-6}$\,ph\,cm$^{-2}$\,s$^{-1}$ (2018 Mar) to $\sim9\times10^{-6}$\,ph\,cm$^{-2}$\,s$^{-1}$ (2022 Sep), with typical uncertainties of $\sim30\%$. They obtained an average centroid energy in 2018--2022 as $\sim6.66$\,keV. Despite the large uncertainties, the centroid energies measured in 2020 and 2022 ($6.63\pm0.09$\,keV and $6.61\pm0.05$\,keV, respectively) seem to be lower than those measured in 2018-2019 ($\sim6.7$\,keV). Nevertheless, no statistically significant variation can be possibly addressed. Their results on Fe K line flux and centroid energy are generally in agreement with ours. The plausible recent decrease in centroid energy has not been revealed by Chandra observations, but should be continuously monitored and further investigated in future works.}

\section{Summary and Conclusions}\label{sec:conclusion}

We performed a comprehensive DEM analysis on the X-ray emitting gas in SN 1987A from 2007 to 2021 based on XMM-Newton RGS $+$ EPIC-pn observations. We obtained the continuous temperature distribution of SN 1987A and followed its up-to-date evolution. Our results make it possible to compare the observed temperature structure directly with those predicted by MHD simulations, and by doing that we found an excellent consistency between observations and simulations. Based on this consistency, we identified {a recent brightening of the reverse shock-heated} ejecta component and confirmed the fading of the ER. The main results and conclusions of this work are summarized as below.

\begin{enumerate}
    \item The X-ray plasma in SN 1987A followed a similar temperature distribution from 2007 to 2021, which consists of a major peak at $\sim0.5$--$1$\,keV and a tail to the high-temperature end (up to $\gtrsim5$\,keV).
    \item The major peak gradually moved from $\sim0.5$\,keV to $\sim1$\,keV, indicating an increase in the average temperature. The total EM of this peak kept climbing in the first few years, reached its maximum at around 2011--2014, and then started to decline. On the other hand, the high-temperature tail shows a continuously increase in its EM and seems to {have} formed a secondary peak at $\sim3$--$5$\,keV in the recent few years.
    \item The DEM results show good consistency with the MHD simulation predictions, in all the following aspects: the temperature and EM of the major peak, the existence of a high-temperature tail and its extent, and the overall evolution of the DEM profile. By comparing observations with simulations, we argue that the recent decline of the major peak and the rapid rising of the tail (the emergence of a secondary peak) reveals the fading of the ER and the brightening of the shocked ejecta.
    \item We {did not find a} recent increase in the average metal abundance of SN 1987A. On the contrary, after a plateau around 2009--2016 ($\sim8000$--11000\,days), the abundances have been gradually decreasing and recently got back to the level as they were around 2007 ($\sim7000$\,days). {We interpret this result as evidence that the forward shock has left the main ER (with higher metal abundances; \citealt{2021ApJ...916...41S}),} while the reverse shock has not yet {reached} the metal-rich inner ejecta.
    \item We found a {moderate} decline of the centroid energy of Fe K line from $\sim6.65$--$6.7$\,keV to $\lesssim6.6$\,keV in the recent few years, {implying} a decrease in the average ionization level of Fe. This is consistent with MHD simulations which predict the Fe line emission to be more and more contributed by the newly shocked {outermost} ejecta with a lower ionization parameter {(and abundances similar to those of the CSM)}, and thus {may} provide additional evidence for the brightening of {the outermost layers of ejecta}.
\end{enumerate}

This work shows again the great importance of SN 1987A as a representative SN/SNR in the sense of linking observations with simulations. We hope the future multiwavelength monitoring of SN 1987A, and the high-resolution X-ray spectroscopic studies using XRISM, HUBS, LEM, etc., will continuously {deepen} our insight into this fascinating object and the underlying physics.

\begin{acknowledgments}

	L.S.\ and Y.C.\ acknowledge the NSFC fundings under grants 12173018, 12121003, and 12393852. L.S.\ acknowledges the support from Jiangsu Funding Program for Excellent Postdoctoral Talent (2023ZB252). P.Z.\ thanks the support from NSFC grant No.\ 12273010.
    S.O., E.G., and M.M. acknowledge financial contribution from the PRIN 2022 (20224MNC5A) - ``Life, death and after-death of massive stars'' funded by European Union – Next Generation EU, and the INAF Theory Grant ``Supernova remnants as probes for the structure and mass-loss history of the progenitor systems''.

\end{acknowledgments}

\software
{XSPEC} \citep{1996ASPC..101...17A}, SPEX \citep{1996uxsa.conf..411K}, SAS \citep{2004ASPC..314..759G}

\appendix

\section{{Selection of the DEM model parameter M and N}}\label{app:M_N}

\begin{figure*}[ht]
    \centering
    \includegraphics[width=1\linewidth]{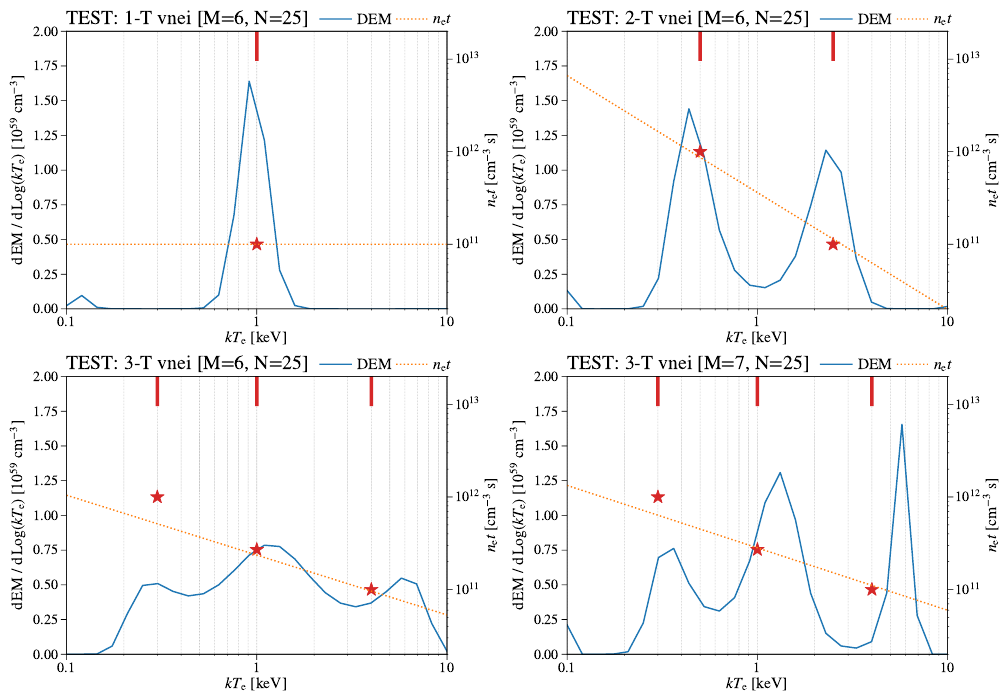}
    \caption{{Similar to Figure \ref{fig:test}, but adopting different $M$, $N$ configurations. The blue solid curves denote the best-fit DEM distributions and the orange dotted lines denote the best-fit ionization parameter distributions. The red lines and stars denote the temperature and ionization parameters used in theoretical models. (a) $M=6$, $N=25$ DEM model fit to 1-T {\tt vnei} test spectra; (b) $M=6$, $n=25$ DEM model fit to 2-T {\tt vnei} test spectra; (c) $M=6$, $N=25$ DEM model fit to 3-T {\tt vnei} test spectra; (d) $M=7$, $N=25$ DEM model fit to 3-T {\tt vnei} test spectra.}}
    \label{fig:dem_test_app}
\end{figure*}

{As mentioned in Section \ref{sec:DEM_model}, the maximum order of the Chebyshev series $M$ and the total temperature bin number $N$ may affect the fitting results. We thereby optimized the selection of $M$ and $N$ by experimenting with various configurations to evaluate their performance. In this section, we present several test cases utilizing different M and N configurations (as shown in Figure \ref{fig:dem_test_app}). For instance, by setting $M=6$ and $N=25$, we found that it yields acceptable fits for the 1-T and 2-T test models (Figure \ref{fig:dem_test_app} a and b). However, it fails to distinctly resolve the individual components in the 3-T test, where the three peaks (correspond to three delta functions in the theoretical model) are conflated with each other (Figure \ref{fig:dem_test_app} c). It also significantly overestimates the temperatures of the 1 keV and 4 keV components, resulted in two peaks around 1.2-1.3 keV and 6-7 keV. Increasing $M$ to 7 improves the fit and helps to better resolve the three components. However, the low- and middle-temperature components remain inadequately separated (Figure \ref{fig:dem_test_app} d), and the overestimation on temperatures remain notable (the high-temperature component is now lying in ~5-6 keV, a little bit closer to its theoretical value comparing with M=6). When adopting $M=7$ and $N=40$ (as illustrated by Figure \ref{fig:test} in Section \ref{sec:test}), all three peaks can be clearly distinguished. the discrepancies between the estimated and theoretical temperatures are now acceptable (up to $\sim10\%$ at 1 keV and up to $\sim25\%$ at 4 keV, which are taken as systematic uncertainties and will not affect our final conclusions). We have also explored even larger values for $M$ and $N$ (e.g., $M=8$ and $N=60$) in pursuit of enhanced resolution capabilities. Nevertheless, the improvements are marginal. Specifically, fine structures such like those in simulated DEMs cannot be fully resolved regardless of how much we increase $M$ and $N$, and the systematic uncertainties of temperature estimation will not be significantly reduced. On the other hand, the computational times for spectral fitting and MCMC running escalate significantly as $M$ and $N$ increasing. As a result, we selected $M=7$ and $N=40$ as the most efficient configurations for our DEM modeling.}

\section{Detailed DEM fitting results and MCMC corner plots}\label{app:corner_plot}

{In this appendix we provide the detailed fitting results (i.e., best-fit parameters and their 1-$\sigma$ errors)  together with the MCMC corner plots {and the fitted spectra} for our DEM modeling.}

\begin{figure*}
    \centering
    \includegraphics[width=0.9\textwidth]{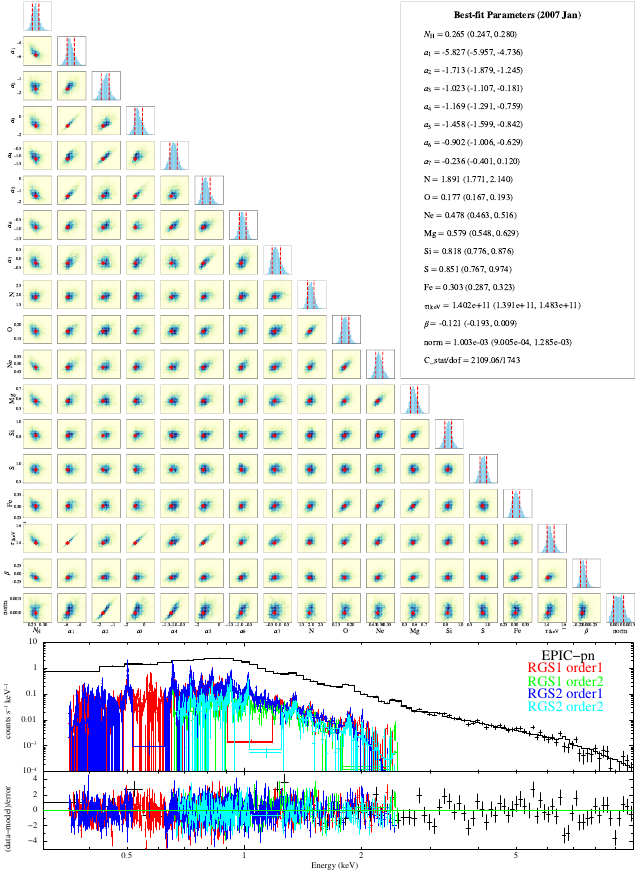}
    \caption{DEM fitting results, MCMC corner plots, {and the fitted spectra} for the observation taken in January 2007. The values of $N_{\rm H}$, $\tau_{\rm 1keV}$, and norm are in units of $10^{22}$\,cm$^{-2}$, cm$^{-3}$\,s, and cm$^{-5}$, respectively.}
    \label{fig:dem_2007_Jan}
\end{figure*}

\clearpage

\begin{figure*}
    \centering
    \includegraphics[width=0.9\textwidth]{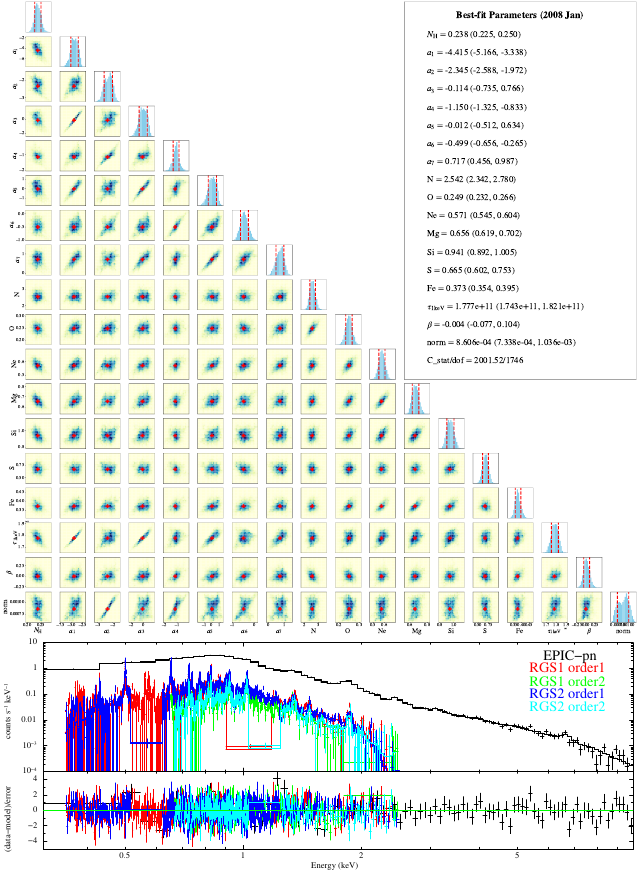}
    \caption{Same as Figure \ref{fig:dem_2007_Jan}, but for the observation taken in January 2008.}
    \label{fig:dem_2008_Jan}
\end{figure*}

\clearpage

\begin{figure*}
    \centering
    \includegraphics[width=0.9\textwidth]{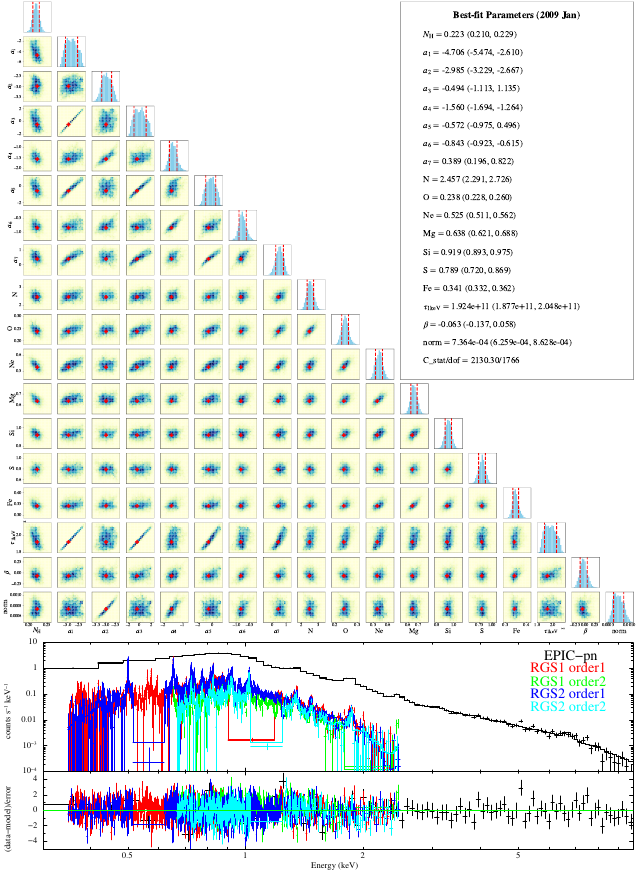}
    \caption{Same as Figure \ref{fig:dem_2007_Jan}, but for the observation taken in January 2009.}
    \label{fig:dem_2009_Jan}
\end{figure*}

\clearpage

\begin{figure*}
    \centering
    \includegraphics[width=0.9\textwidth]{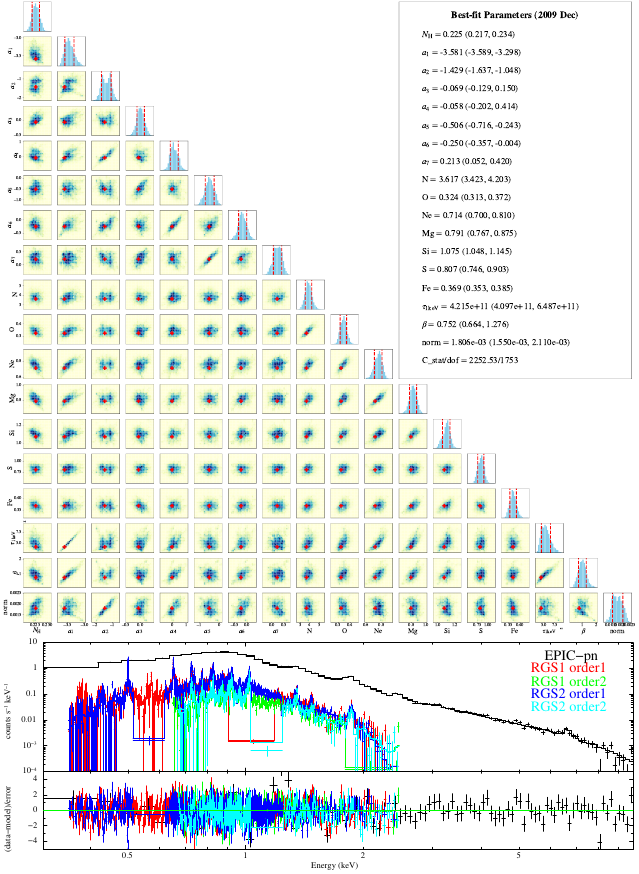}
    \caption{Same as Figure \ref{fig:dem_2007_Jan}, but for the observation taken in December 2009.}
    \label{fig:dem_2009_Dec}
\end{figure*}

\clearpage

\begin{figure*}
    \centering
    \includegraphics[width=0.9\textwidth]{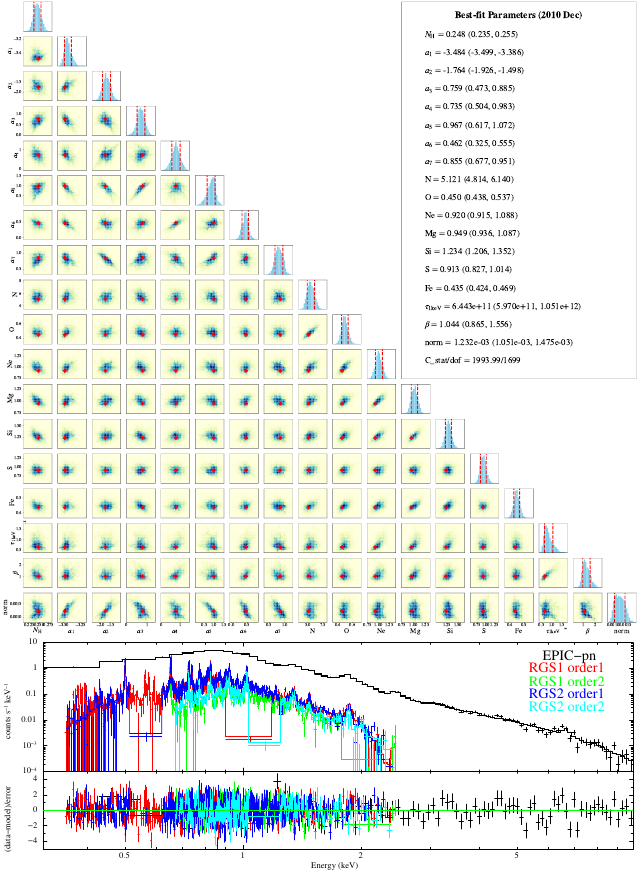}
    \caption{Same as Figure \ref{fig:dem_2007_Jan}, but for the observation taken in December 2010.}
    \label{fig:dem_2010_Dec}
\end{figure*}

\clearpage

\begin{figure*}
    \centering
    \includegraphics[width=0.9\textwidth]{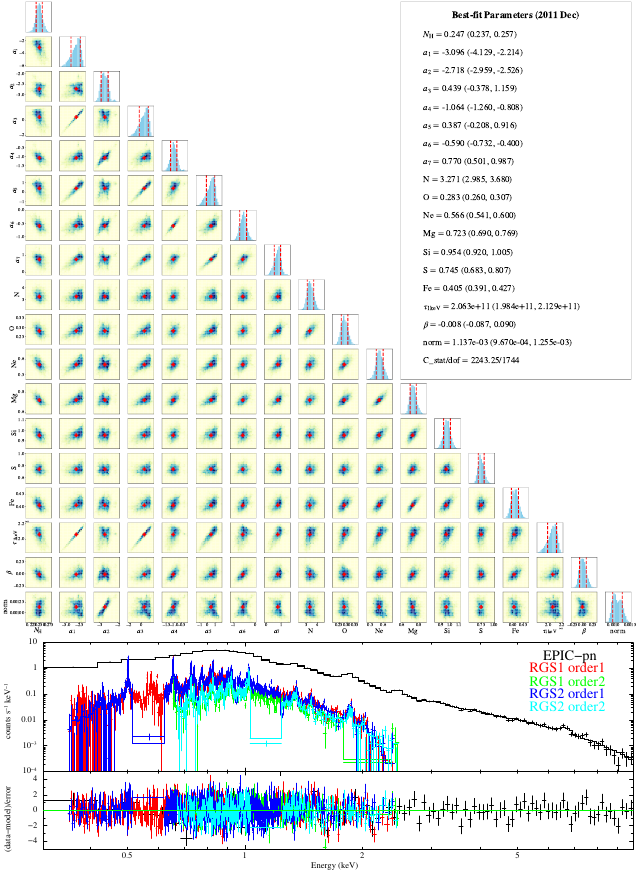}
    \caption{Same as Figure \ref{fig:dem_2007_Jan}, but for the observation taken in December 2011.}
    \label{fig:dem_2011_Dec}
\end{figure*}

\clearpage

\begin{figure*}
    \centering
    \includegraphics[width=0.9\textwidth]{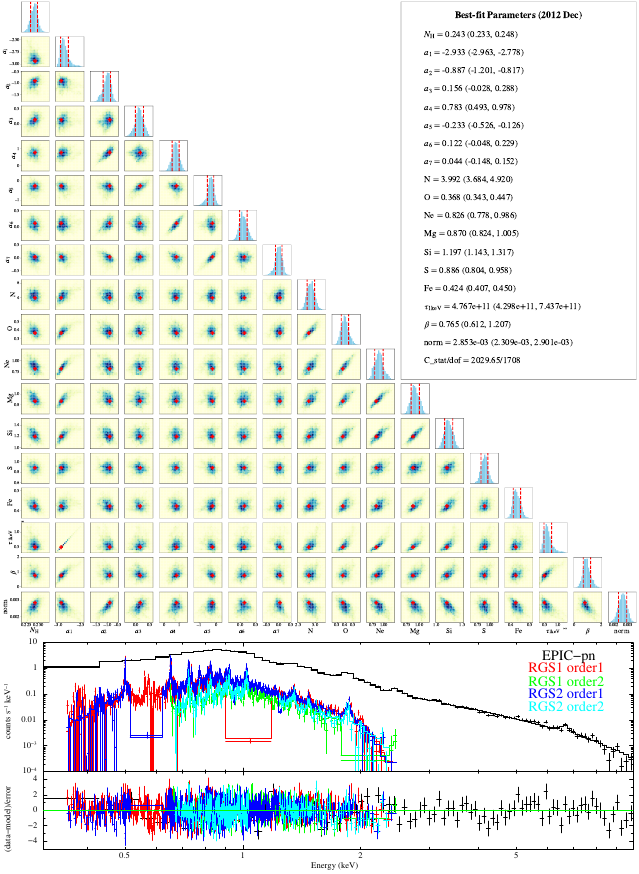}
    \caption{Same as Figure \ref{fig:dem_2007_Jan}, but for the observation taken in December 2012.}
    \label{fig:dem_2012_Dec}
\end{figure*}

\clearpage

\begin{figure*}
    \centering
    \includegraphics[width=0.9\textwidth]{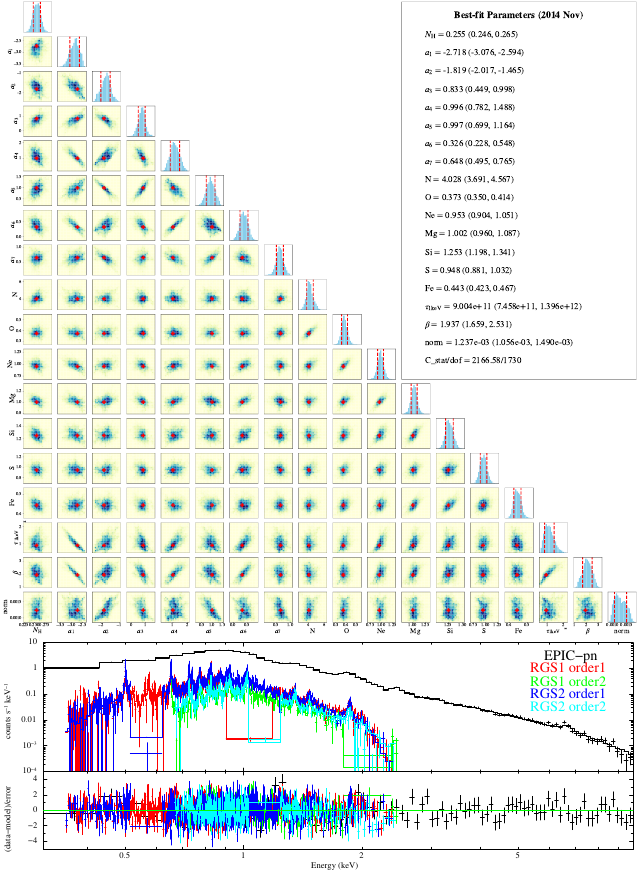}
    \caption{Same as Figure \ref{fig:dem_2007_Jan}, but for the observation taken in November 2014.}
    \label{fig:dem_2014_Nov}
\end{figure*}

\clearpage

\begin{figure*}
    \centering
    \includegraphics[width=0.9\textwidth]{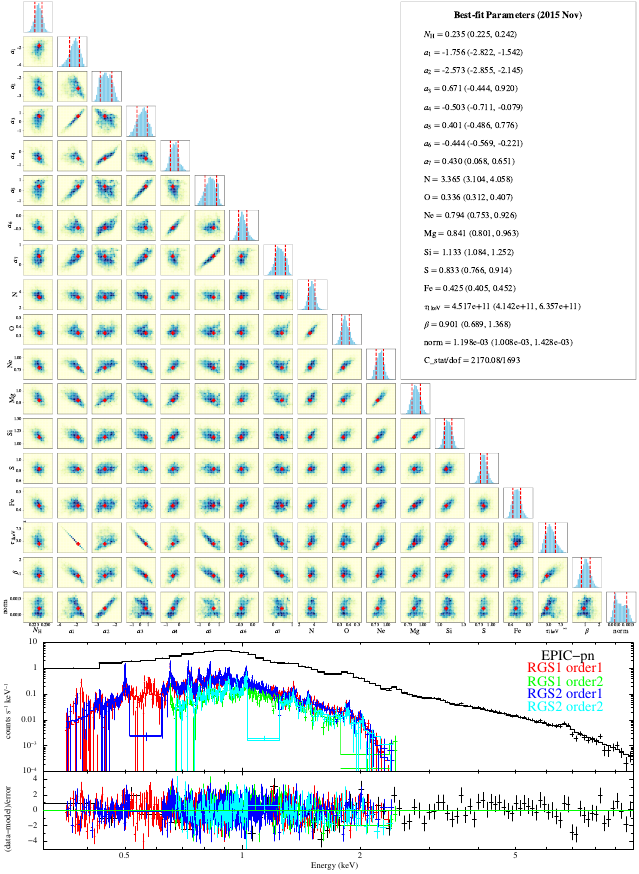}
    \caption{Same as Figure \ref{fig:dem_2007_Jan}, but for the observation taken in November 2015.}
    \label{fig:dem_2015_Nov}
\end{figure*}

\clearpage

\begin{figure*}
    \centering
    \includegraphics[width=0.9\textwidth]{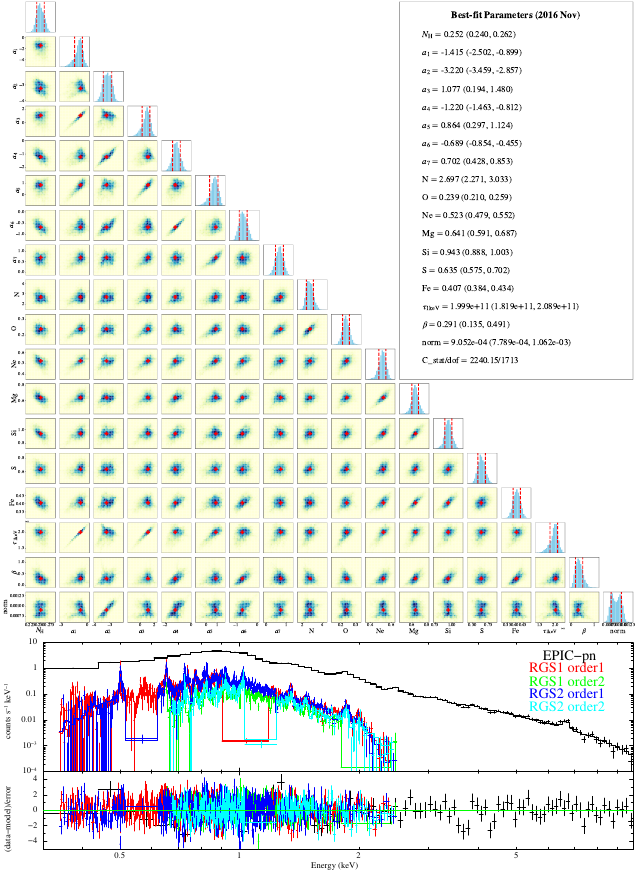}
    \caption{Same as Figure \ref{fig:dem_2007_Jan}, but for the observation taken in November 2016.}
    \label{fig:dem_2016_Nov}
\end{figure*}

\clearpage

\begin{figure*}
    \centering
    \includegraphics[width=0.9\textwidth]{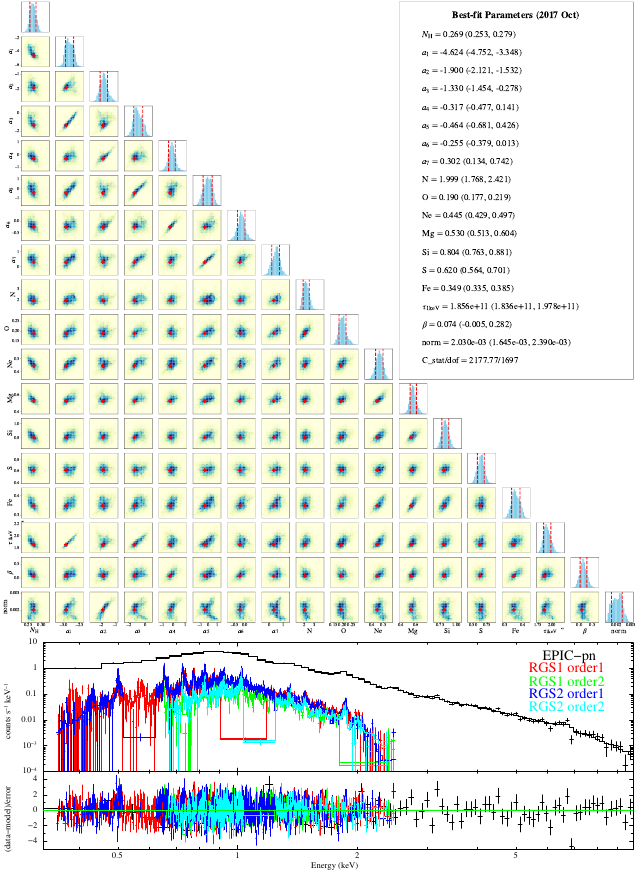}
    \caption{Same as Figure \ref{fig:dem_2007_Jan}, but for the observation taken in October 2017.}
    \label{fig:dem_2017_Oct}
\end{figure*}

\clearpage

\begin{figure*}
    \centering
    \includegraphics[width=0.9\textwidth]{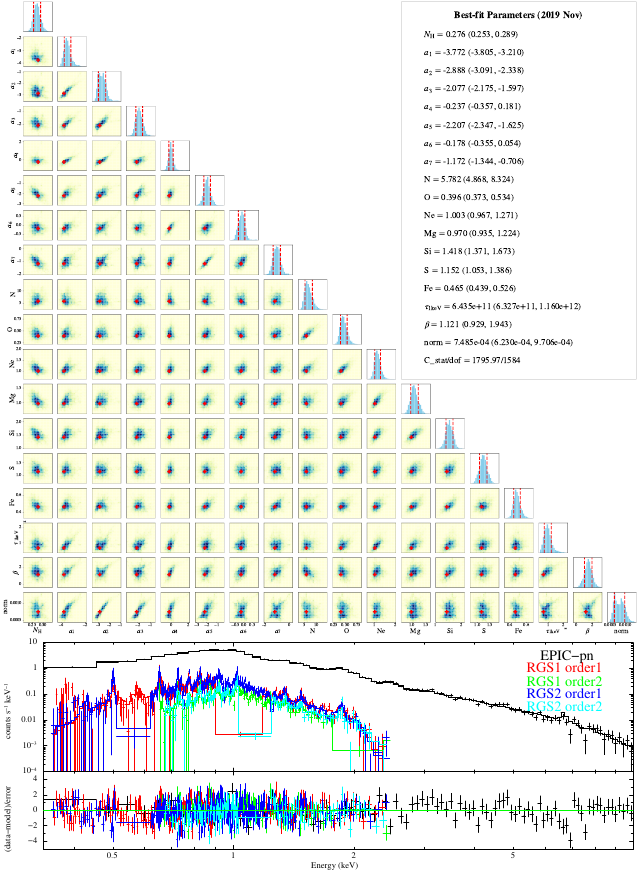}
    \caption{Same as Figure \ref{fig:dem_2007_Jan}, but for the observation taken in November 2019.}
    \label{fig:dem_2019_Nov}
\end{figure*}

\clearpage

\begin{figure*}
    \centering
    \includegraphics[width=0.9\textwidth]{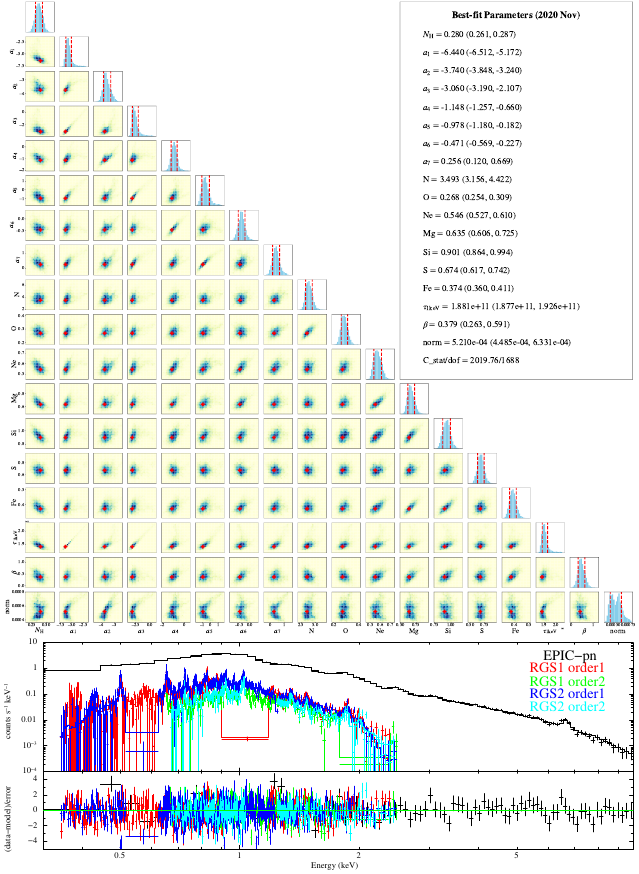}
    \caption{Same as Figure \ref{fig:dem_2007_Jan}, but for the observation taken in November 2020.}
    \label{fig:dem_2020_Nov}
\end{figure*}

\clearpage

\begin{figure*}
    \centering
    \includegraphics[width=0.9\textwidth]{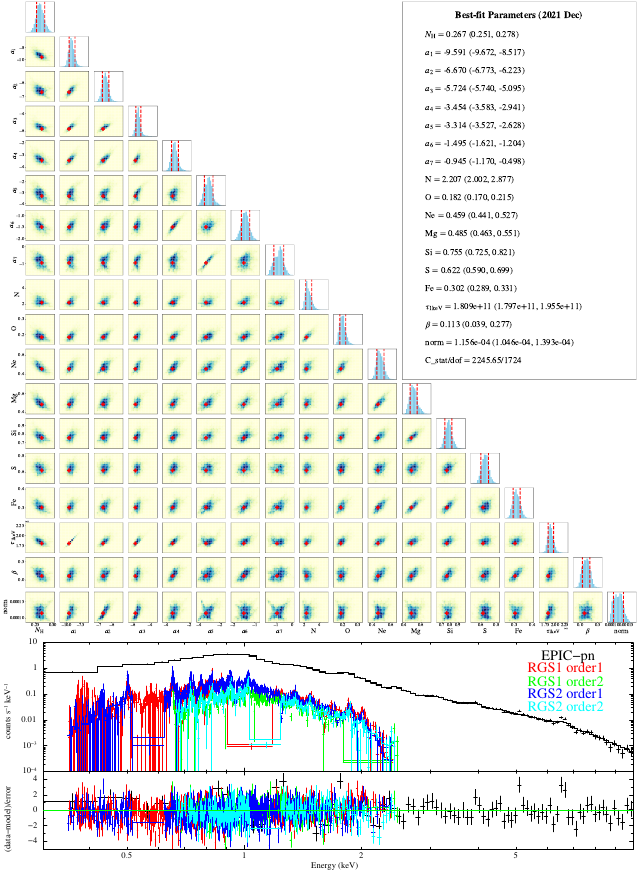}
    \caption{Same as Figure \ref{fig:dem_2007_Jan}, but for the observation taken in December 2021.}
    \label{fig:dem_2021_Dec}
\end{figure*}

\clearpage

\section{DEM analysis results based on the {\tt vpshock} model} \label{apx:vpshock}

\begin{figure*}
    \centering
    \includegraphics[width=\textwidth]{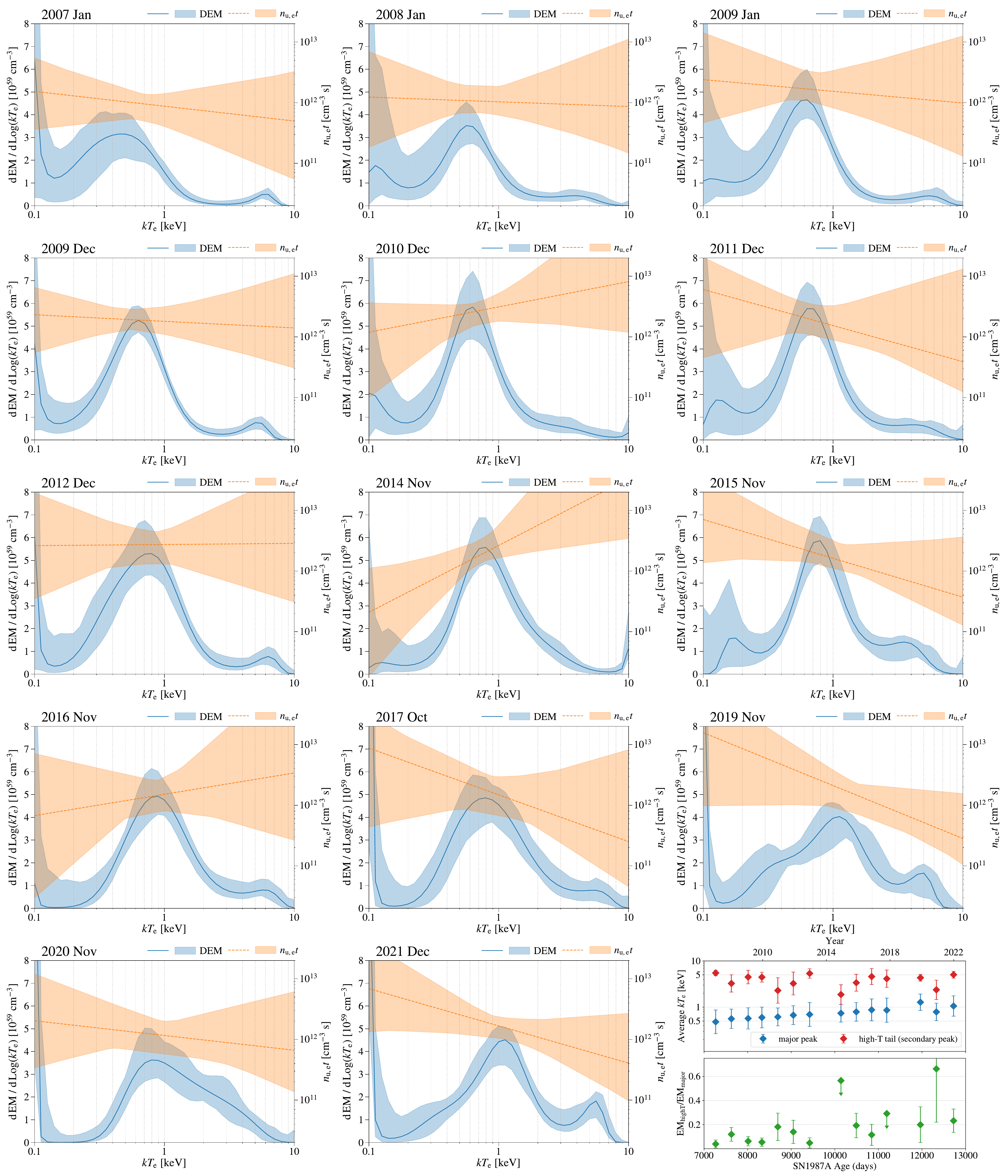}
    \caption{DEM fitting results for SN 1987A based on the {\tt vpshock} model.}
    \label{fig:dem_vpshock}
\end{figure*}

{Despite the disadvantages with respect to {\tt vnei}-based DEM modeling as illustrated in Section \ref{sec:compare}, we constructed a DEM model based on the {\tt vpshock} model in XSPEC analogous to the {\tt vnei}-based model as described in Section \ref{sec:DEM_model}, in order to compare their results. We applied this {\tt vpshock}-based DEM model to the RGS + EPIC-pn data of SN 1987A. The fitting results (i.e., the best-fit DEM and $n_{\rm e}t$ with their 1-$\sigma$ errors as a function of $kT_{\rm e}$) are shown in Figure \ref{fig:dem_vpshock}. Similar to the {\tt vnei}-based fitting, {\tt vpshock}-based DEM modeling also revealed a major peak and a high-temperature tail in the DEM distribution of SN 1987A, and captured the temporal evolution of their {average temperatures and} EMs. {The average temperature of the major peak guadually increased from $\sim0.5$\,keV to $\sim1$\,keV.} A recent rising of the high-temperature tail can also be seen in the {\tt vpshock}-based fitting, {resulting in a increase in the EM ratio ${\rm EM}_{\rm highT}/{\rm EM}_{\rm major}$}. 

The most prominent difference between the {\tt vpshock}-based and {\tt vnei}-based DEM modeling is in the obtained ionization parameter distributions. The ionization parameters obtained by the {\tt vpshock}-based modeling appear to be generally higher than those obtained by {\tt vnei}-based modeling, which is because that the $n_{\rm u,e}t$ in {\tt vpshock} model is defined as the upper limit of the ionization parameter rather than the average value $n_{\rm e}t$ in {\tt vnei}. In contrast with the average $n_{\rm e}t$ given by {\tt vnei}-based fitting showing a positive relation with temperature, $n_{\rm u,e}t$ given by {\tt vpshock}-based fitting exhibits a flat or negative relation with $kT_{\rm e}$ (i.e., $\beta\lesssim0$ in Eq. \ref{eq:net}). Despite the rather large uncertainties, this may indicate that for the high-temperature gas, the average ionization parameter is closer to its upper limit, while for the low-temperature gas, even though the upper limit of the ionization parameter is rather high, its average value is still low. This is consistent with the findings of \citet{2021ApJ...916...41S} based on 2-T {\tt vnei} and {\tt vpshock} fitting, and might imply the different ionization histories of the high- and low-temperature gas.}

\section{{Statistical significance of the recent decrease in Fe K centroid energy}}\label{apx:Fe_K}

\begin{figure*}[ht]
    \centering
    \includegraphics[width=0.49\linewidth]{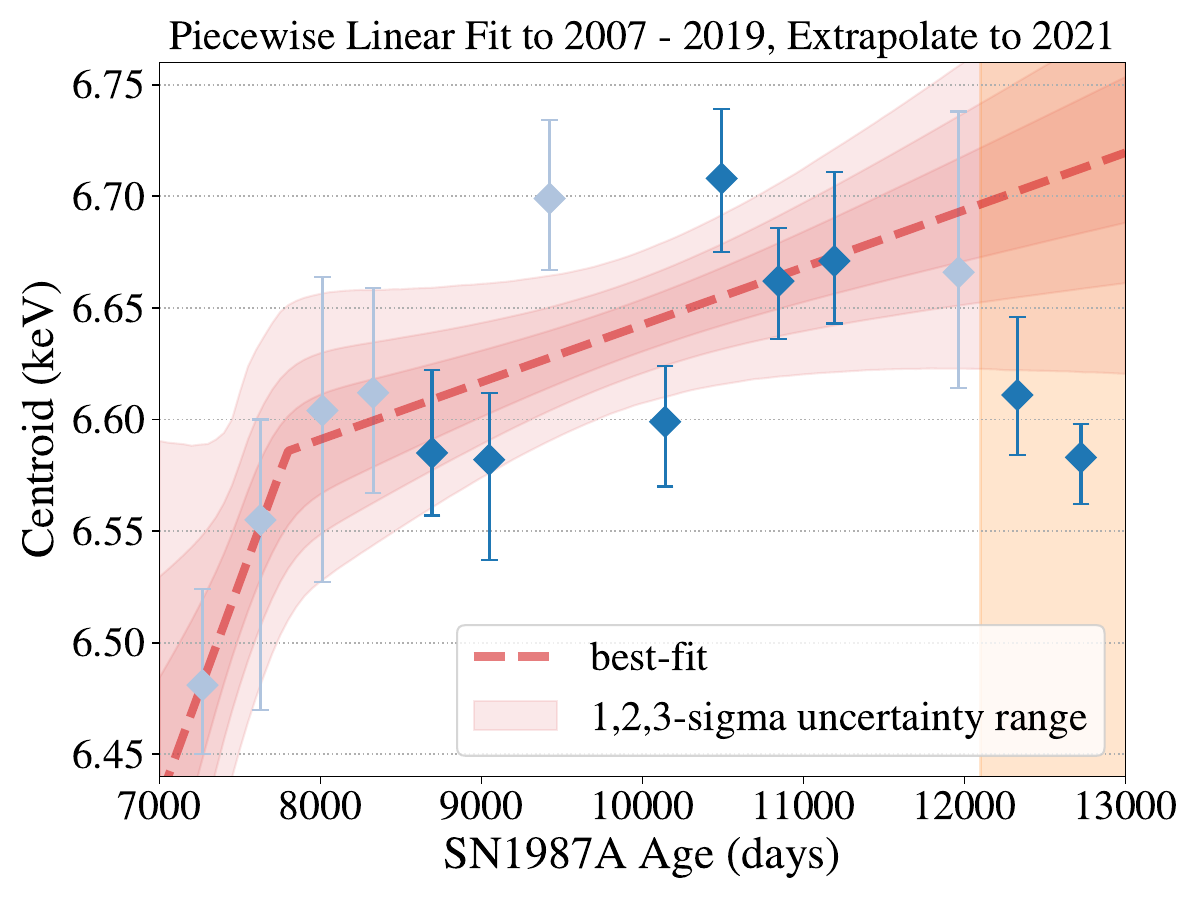}
    \includegraphics[width=0.49\linewidth]{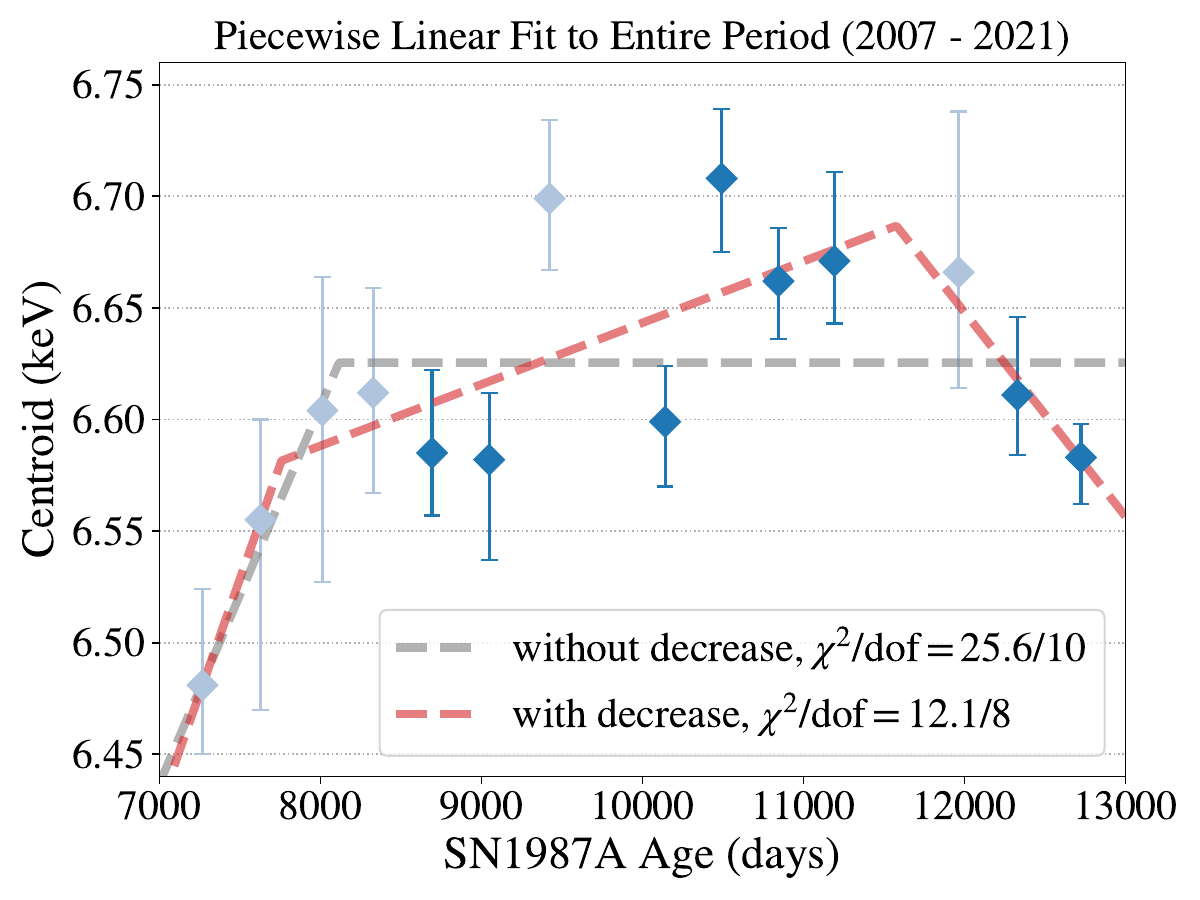}
    \caption{Piecewise linear fit to the Fe K centroid energy. Left: fit to the observations from 2007 to 2019, and then extrapolate the result to 2021. The red dashed line denotes the best-fit model, and the shaded regions indicate the 1,2,3-sigma uncertainty ranges. Right: fit to the entire period (2007 to 2021). The gray dashed line denotes the best-fit model without a recent decrease, while the red dashed line denotes the best-fit model with a recent decrease.}
    \label{fig:Fe_K_significance}
\end{figure*}

{
In order to approximately evaluate the statistical significance of the recent decrease in centroid energy of Fe K line, we followed the subsequent procedures. 

First, we assumed that the centroid energy remains constant (at the average value) throughout the entire period (2007--2021) and calculated the chi-squared. We obtained the $\chi^2/{\rm dof}\sim40.8/13$, corresponding to p-value\,$\sim1\times10^{-4}$, suggesting that the centroid energy is not likely to be constant throughout the entire period. 

Subsequently, we fitted the observed centroid energies from 2007 to 2019 (excluded the last two data points) using a piecewise linear function with two segments. This model provides a good fit to the Fe K centroid energy from 2007 to 2019, with a $\chi^2/{\rm dof}\sim12.1/8$. The best-fit result is presented as the red dashed curves in the left panel of the Figure \ref{fig:Fe_K_significance} (the 1,2,3-sigma uncertainty ranges are denoted by the shaded regions). The fitting result indicates a fast increase of Fe K centroid energy before $\sim8000$\,days after the explosion, following by a slower increase until $\sim12000$\,days. If we extrapolate the best-fit curve and the uncertainty ranges to 2021, we found that the two new observations show a significant deviation from the best-fit, which fall out of the 3-sigma region, indicating a recent decrease in Fe K centroid energy. However, we noted that some of the 2007--2019 observations also show large scatterings from the best-fit curve, lying outside the 3-sigma region (e.g., 2012, 2014, and 2015).

At last, we fitted the observations in the entire period (2007--2021, including the last two data points) with piecewise linear functions. We firstly assumed that there is no decrease --- the centroid energy increased at beginning, and then keep constant after some point. The best-fit result in this case (without decrease) is shown as the gray dashed line in the right panel of Figure \ref{fig:Fe_K_significance}, which gives a $\chi^2/{\rm dof}\sim25.6/10$. Then, we introduced a recent decrease in to the fitting model. The best-fit result in this case (with decrease) is shown as the red dashed line, which gives a $\chi^2/{\rm dof}\sim12.1/8$. An f-test between the two models gives p-value\,$\sim0.0497$, which suggests that the with-decrease model to be favored at a $\sim2$-sigma significance level.

Given the investigations above, we considered the recent decrease in Fe K centroid energy to be moderately significant ($\sim2$--$3$ sigma level). 
}

\bibliography{LS_ref}{}
\bibliographystyle{aasjournal}



\end{document}